\PassOptionsToPackage{hyphens}{url}

\documentclass[sigconf]{acmart}
\AtBeginDocument{%
  \providecommand\BibTeX{{%
    \normalfont B\kern-0.5em{\scshape i\kern-0.25em b}\kern-0.8em\TeX}}}

\copyrightyear{2025}
\acmYear{2025}
\setcopyright{acmlicensed}\acmConference[ASSETS '25]{The 27th International ACM SIGACCESS Conference on Computers and Accessibility}{October 26--29, 2025}{Denver, CO, USA}
\acmBooktitle{The 27th International ACM SIGACCESS Conference on Computers and Accessibility (ASSETS '25), October 26--29, 2025, Denver, CO, USA}
\acmDOI{10.1145/3663547.3746466}
\acmISBN{979-8-4007-0676-9/2025/10}
\usepackage{url} 
\usepackage{hyperref}
\usepackage[normalem]{ulem}
\usepackage{multirow}
\usepackage{array}
\usepackage{graphicx}
\usepackage{tabularx}
\usepackage{appendix}
\usepackage[utf8]{inputenc}
\usepackage{pdflscape}
\usepackage{lscape}    % 只旋转内容，不给 PDF 页面设置 Rotate
\usepackage{rotating}   % 提供 sidewaystable
\usepackage{multirow}
\usepackage{longtable}   % 支持跨页表格

\usepackage{ragged2e}    % 支持 \raggedright
\usepackage{caption}     % 如需调整标题样式
\begin{document}

\title[Accessibility and Social Inclusivity]{Accessibility and Social Inclusivity: A Literature Review of Music Technology for Blind and Low Vision People}


\author{Shumeng Zhang}
\affiliation{%
  \institution{The Hong Kong University of Science and Technology (Guangzhou)}
   \city{Guangzhou}
    \state{Guangdong}
  \country{China}}
\email{szhang390@connect.hkust-gz.edu.cn}

\author{Raul Masu}
\affiliation{%
  \institution{Conservatorio di musica F. A. Bonporti di Trento e Riva del Garda}
     \city{Trento}
  \country{Italy}}
  \affiliation{%
 \institution{The Hong Kong University of Science and Technology (Guangzhou)}
 \city{Guangzhou}
 \state{Guangdong}
  \country{China}}
\email{raul.masu@conservatorio.tn.it}

\author{Mela Bettega}
\affiliation{%
  \institution{Newcastle University}
   \city{Newcastle upon Tyne}
  \country{United Kingdom}}
\email{mela.bettega@newcastle.ac.uk}

\author{Mingming Fan}
\authornote{Corresponding author}
\affiliation{%
  \institution{The Hong Kong University of Science and Technology (Guangzhou)}
     \city{Guangzhou}
    \state{Guangdong}
  \country{China}}
  \affiliation{%
 \institution{The Hong Kong University of Science and Technology}
 \city{Hong Kong SAR}
  \country{China}}
\email{mingmingfan@ust.hk}


\renewcommand{\shortauthors}{Zhang et al.}
%\engquote{sfsfsfsfs}
%%

\begin{abstract}
This paper presents a systematic literature review of music technology tailored for blind and low vision (BLV) individuals. Music activities can be particularly beneficial for BLV people. However, a systematic approach to organizing knowledge on designing accessible technology for BLV people has yet to be attempted. We categorize the existing studies based on the type of technology and the extent of BLV people's involvement in the research. We identify six main categories of BLV people-oriented music technology and highlight four key trends in design goals. Based on these categories, we propose four general insights focusing on (1) spatial awareness, (2) access to information, (3) (non-verbal) communication, and (4) memory. The identified trends suggest that more empirical studies involving BLV people in real-world scenarios are needed to ensure that technological advancements can enhance musical experiences and social inclusion. This research proposes collaborative music technology and inclusive real-world testing with the target group as two key areas missing in current research. They serve as a foundational step in shifting the focus from ``accessible technology'' to ``inclusive technology'' for BLV individuals within the broader field of accessibility research.  
\end{abstract}


%handicrafts

%   \includegraphics[width=\textwidth]{main/teaser.jpg}
%   \caption{Beadwork created by BLV students. (a) A plane bracelet beadwork. (b) A plane cardholder beadwork. (c) A symmetrical 3D monkey beadwork. (d) A distorted monkey beadwork. (e) A series of vases with the color and pattern changing. (f) A symmetrical 3D traditional house beadwork. (g) An asymmetrical 3D beadwork. (h) A beadwork incorporates additional handicraft materials. Moreover, these figures show the progression from the simple to the difficult that teachers follow when teaching beadwork.}
%   \Description{The Beadwork created by BLV students. (a) A plane bracelet beadwork. (b) A plane cardholder beadwork. (c) A symmetrical 3D monkey beadwork. (d) A distorted monkey beadwork. (e) A series of vases with the color and pattern changing. (f) A symmetrical 3D traditional house beadwork. (g) An asymmetrical 3D beadwork. (h) A beadwork incorporates additional handicraft materials. Moreover, these figures show the progression from the simple to the difficult that teachers follow when teaching beadwork.}
%   \label{fig:bw}



\maketitle

\section{Introduction}
Music enriches daily life, fostering self-expression and social engagement \cite{hoppestad2007inadequacies,mack2021we}. Additionally, music can help people with disability to build a sense of community and social connections \cite{veblen2002community,koopman2007community,baker2018disability,boer2009music,mcdermott2014importance}. For individuals who are Blind and Low Vision (BLV), music practice can be particularly fruitful. Indeed, BLV individuals have been active participants in various musical scenes, and a long history of BLV musicians showcase how music can help BLV people find a respected and professionally recognized place within society\cite{milburn2018blind,kononenko2015ukrainian,de2010last,streatfield2018revival}. 
Additionally, cognitive psychology evidence suggests that BLV individuals often exhibit aural perception abilities above the average \cite{mitchell2017teaching,niemeyer1981blind,sabourin2022blind}. 
Overall, music can be useful for BLV people both individually (developing skills) and socially (engaging with others and society). 
For this reason, it also fits well within the recent shift from understanding disability through individual actions toward understanding it within its social frame \cite{hoppestad2007inadequacies,mack2021we,trevisan2022beyond,jian2020spatial}.

These elements underscore the importance of designing inclusive music technologies for BLV people. 
However, music technology for BLV people remains notably under-studied. 
While many papers have systematically analyzed different technologies for BLV people - e.g., mobile platforms \cite{csapo2015survey}, internet access \cite{perfect2019systematic}, and navigation tools \cite{kuriakose2022tools}, a 2019 systematic review of accessible music technology in general (for all different populations) highlighted the lack of studies focusing on accessible music technology for BLV people \cite{dd2019accessible}.  
The scarce research drives us to systematically scrutinize the current landscape of the research on music technology for BLV individuals, in order to identify trends, gaps, insights, and point out future directions. 

\begin{figure*}[ht]
\centering
\includegraphics[width=0.9\textwidth]{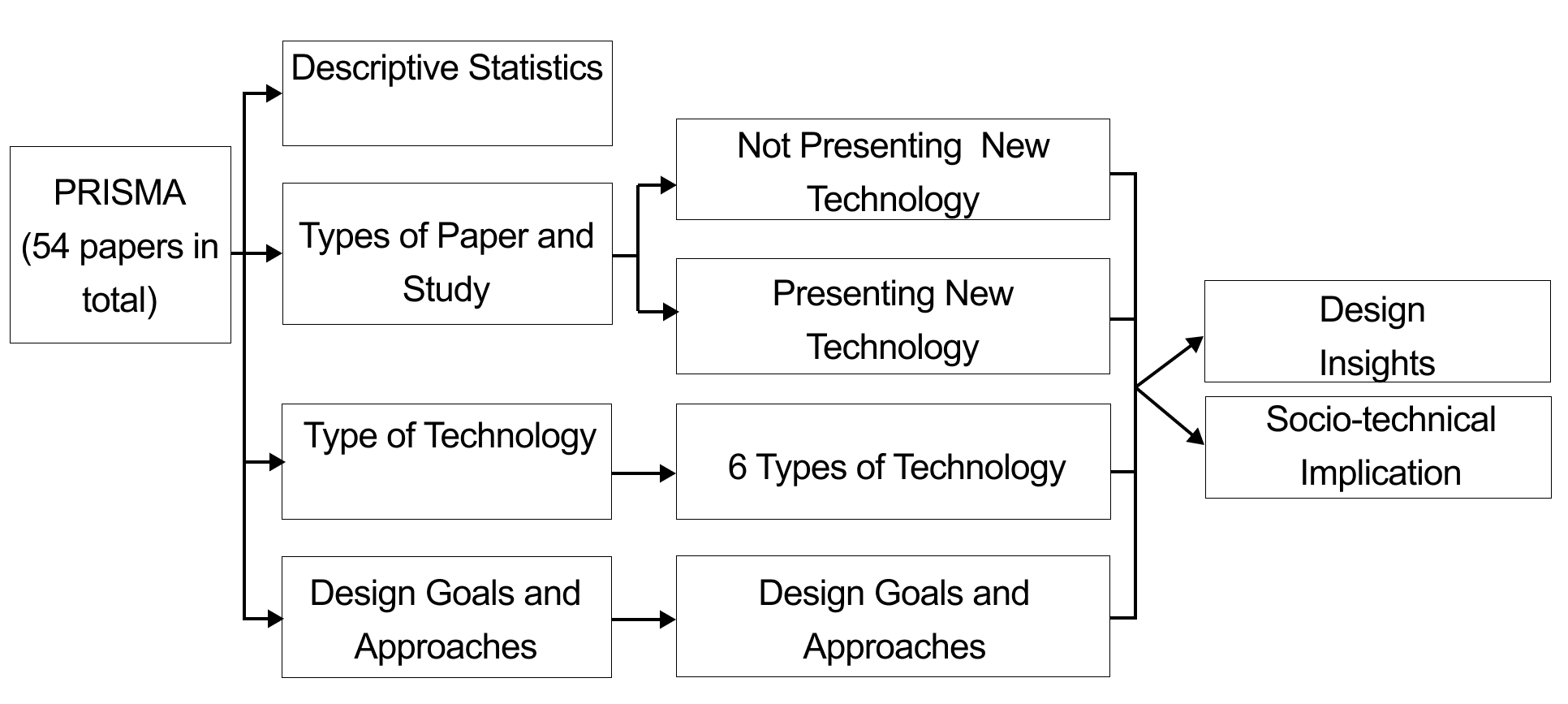}
\caption{\textbf{An overview of the structure of the paper, outlining how we analyzed the studies and how we built the discussion from our analysis.}}
\label{fig:intro}
\Description{This figure illustrates the overview of the structure of the paper, outlining how we analyzed the studies and how we built the discussion from our analysis. In this figure, we outline the structure of a paper based on 42 PRISMA-reviewed studies. It covers descriptive statistics, 2 study types, and 6 types of technologies, and 4 current trends of design goals and approaches. The review further leads to 4 design insights, and future direction on socio-technical implications.}
\end{figure*}

Informed by this gap, this paper presents a systematic literature review of music technology for BLV people. 
As Figure \ref{fig:intro} shows, we adopted Preferred Reporting Items for Systematic Reviews and Meta-Analyses (PRISMA) method \footnote{See: \url{https://www.prisma-statement.org/}}, and identified a corpus of 54 papers. 
After analyzing descriptive trends on publications per year and venues, we clustered these studies based on two criteria 
1) \textit{type of studies} - 
taking into consideration the level of engagement of BLV people in the study; 
and 2) \textit{type of technology} - for what type of music activity the technology was designed for (more details in section \ref{sec:methdology}). 

We identified two main types of research: research presenting new technology; and research not presenting new technology (e.g. field work or position papers). 
For the papers presenting new technology, we also identified different levels of BLV people engagement in the research (from no engagement to co-design) and highlighted the scarcity of studies investigating these technologies in real-world contexts. 
Our second analysis further identified six types of music technologies tailored to BLV individuals (results outlined in section \ref{sec:Res}): 1) Technology for Reading Music, 
2) Technology for Learning Traditional Music, 
3) Technology for Studio Activities, 
4) New Digital Musical Instruments, 
5) Music Technology for Non-musical Skills Acquisition,
and 6) Music Technology and Video Games. 
Finally, we concluded the paper with four design goals and approaches \ref{sec:goals} from the existing design trends and challenges in the papers we reviewing: 
(1) fostering spatial awareness, 
(2) supporting intuitive access to information, 
(3) facilitating non-verbal communication, 
and (4) enhancing physical memory.

With alignment of the current design goals and approaches, the paper continues with discussing four design insights (section \ref{sec:discussion}): 
(1) Fostering Spatial Awareness by Reducing Cognitive Overloads;
(2) Accessing Music with Alternative Music Representation;
(3) Supporting Memory by Embodied Interactions;
(4) Facilitating  (Non-verbal) Collaboration with Multi-sensory Feedback.

In conclusion (section \ref{sec:socio}), we reflected on the methodologies employed and argue for the need to bridge the gap from accessibility to social inclusivity in music technology design for BLV individuals. This requires longer-term studies that engage with real-world contexts.
Overall, the paper makes two main contributions to the following aspects: 
1) The paper provides a classification of existing paper types, technologies for different BLV people's music activities, and design goals and approaches. Through this classification, the paper reveals the current trends in music technologies and design goals.
2) The paper highlights a gap in existing research on music technology for BLV people, specifically the lack of testing with target participants in real-world contexts. This presents an opportunity for future research to focus on design goals and approaches that promote social inclusivity. 
3) This research proposes collaborative music technology and inclusive real-world testing with the target group as two key areas missing in current research. They serve as a foundational step in shifting the focus from ``accessible technology'' to ``inclusive technology'' for BLV individuals within the broader field of accessibility research.
\section{Background}
\subsection{Music and BLV people}
\label{2.1}
% As mentioned in the introduction, engaging in music practice offers a range of benefits, from personal self-satisfaction to enhanced social activity \cite{north2000importance,boer2009music,mcdermott2014importance,baker2017insights}. Additionally, recent literature has underscored how music-making can aid people with disabilities in personal development as well as connecting with their communities and ultimately fostering social connections \cite{veblen2002community,koopman2007community,baker2018disability,boer2009music,mcdermott2014importance}. Music holds significant importance for BLV individuals, as evidenced by numerous historical examples of prominent blind musicians across cultures. In the West, trends of blind musicians have been identified toward a specific instrument (i.e., the organ \footnote{\url{https://www.voxhumanajournal.com/langlais2021.html}}, or a specific social practice (i.e., itinerary musicians in Ukraine \cite{kononenko2015ukrainian}). In Asia, Shi Kuang, a blind musician from the 6th century BC, is among the earliest known performers, and China's Guilds of Blind Musicians and Fortune-Tellers remained active until the mid-20th century \cite{milburn2018blind}. Japan's biwa hoshi, traveling musicians, were also frequently blind \cite{de2010last}. 
Engaging in music practice offers benefits from personal satisfaction to enhanced social activity \cite{north2000importance,boer2009music,mcdermott2014importance,baker2017insights}. Recent studies show that music-making helps people with disabilities in personal development and community connection \cite{veblen2002community,koopman2007community,baker2018disability,boer2009music,mcdermott2014importance}. Music holds significant value for BLV individuals, as seen in historical examples of blind musicians across cultures. In the West, blind musicians often specialize in specific instruments or social practices \cite{kononenko2015ukrainian}. In Asia, figures like Shi Kuang and blind musicians in China and Japan show the deep-rooted tradition \cite{milburn2018blind,de2010last}. For a broader survey of blind musicians, see \cite{park2017finding,hash2015music}. 

Research shows that integrating music into the lives of BLV people enhances financial independence and enjoyment \cite{baker2021additional,baker2017insights,baker2023disability}. Studies in pedagogy emphasize music education for BLV individuals \cite{baker2016perceptions,abramo2013ethnographic}. Cognitive psychology highlights heightened auditory abilities in BLV individuals, including superior speech discrimination and enhanced detection of musical sound features \cite{mitchell2017teaching,niemeyer1981blind,sabourin2022blind}.

% Existing research in pedagogy has demonstrated that the integration of music into BLV people’s daily life and education can effectively enhance their financial independence and overall enjoyment \cite{baker2021additional,baker2017insights,baker2023disability}, thus the pedagogical exploration of music teaching is a common focal point in academic discourse \cite{baker2016perceptions,abramo2013ethnographic}. While historical evidence and extant studies underscore the pivotal role of music in the lives of BLV individuals, research in cognitive psychology posits that visually impaired individuals exhibit heightened auditory capabilities \cite{mitchell2017teaching,niemeyer1981blind,sabourin2022blind}. For example, research suggests that BLV people have superior speech discrimination abilities \cite{niemeyer1981blind}. Moreover, in a recent meta-analysis of over 100 studies, Sabourin and colleagues highlighted numerous studies showing that BLV people have heightened detection abilities and discrimination capabilities of many sound features relevant to music \cite{sabourin2022blind}. 

To conclude, we wish to emphasize the importance of facilitating access to music-making for BLV people, both as a tool for personal development — with potential pathways to professional engagement — and as a means of fostering social connections. Contemporary technology has been used in different ways, and we are going to analysis in depth all the types of research on music technology in the main corpus of this paper. 

\subsection{Assistive and Inclusive Technology}
Technology for BLV individuals falls under the broader scope of accessible technology. In general, accessibility and inclusivity in Human-Computer Interaction (HCI) focus on designing for people with needs that differ from the average, particularly those with disabilities \cite{mack2021we} (e.g., sensory \cite{jiang2024designing,de2024caption,kim2022participatory}, motor \cite{ascari2020computer,nowrin2022exploring}, or cognitive \cite{chinn2017easy,valencia2019impact} impairments).

Historically, disability was viewed as a medical issue, addressed primarily through individual behaviors or the use of assistive devices \cite{siebers2008disability}. 
However, there has been a recent shift - driven by the disability rights movement \footnote{\url{https://www.huffpost.com/entry/the-global-disability-rig_b_5651235}} - away from this individual-centered perspective toward understanding disability as a social construct. This shift highlights the need for societal change instead of placing the responsibility for adaptation solely on the individual.\cite{hoppestad2007inadequacies,mack2021we,trevisan2022beyond,jian2020spatial}. 
For instance, technology should not merely facilitate disabled people's access to digital content \cite{trevisan2022beyond} or public spaces \cite{jian2020spatial} but rather supporting their active involvement. To evaluate assistive technologies for BLV users, many studies emphasize long-term testing as essential for understanding both self-development and social impact \cite{de2025sensing,adams2016blind,mathis2025lifeinsight,rector2017design}. This work falls within HCI research, where longitudinal approaches are widely recommended for in-context use evaluation \cite{minton2025longitudinal,wang2025facilitating,ambe2022collaborative}.

In the previous Section \ref{2.1}, music expands social networks for BLV people, fosters a sense of belonging, and can open income-generating roles that enhance independence and public visibility \cite{north2000importance,boer2009music,baker2018disability,baker2021additional}. At the same time, instrumental study refines fine-motor control and posture, leverages heightened auditory acuity, and further sharpens sound discrimination skills, thereby promoting confidence and lifelong self-development \cite{niemeyer1981blind,mitchell2017teaching,sabourin2022blind}. Therefore, in our analysis, we will account for both individual access to music and the social inclusivity of music making. 

\subsection{Technology for BLV people}
\label{bk_tecBLV}

A recent literature review on accessibility research \cite{mack2021we} highlights that nearly half of the efforts focus on BLV individuals. One review identified three main technological solutions for BLV people: assistive technology (for daily life), digital access tools (for digital information), and virtual interfaces (for interacting with the world) \cite{ashraf2016systematic}. Another review uncovered key themes in accessible technology research for BLV, such as visual information interpretation, digital content accessibility, mobility, environmental awareness, and multimodal sensory experiences \cite{bhowmick2017insight}. Other reviews focused on mobile platforms \cite{csapo2015survey}, internet access \cite{perfect2019systematic}, diagram accessibility \cite{torres2019approaches}, VR technologies \cite{ghali2012virtual}, and navigation tools \cite{kuriakose2022tools}, with an emphasis on reading and navigation \cite{berla1977tactual,capovilla2013teaching}.

General recommendations for technology development for BLV people include \textit{spatial awareness} and \textit{embodied cognition}. \textit{Spatial awareness} is the ability to perceive one's surroundings and position \cite{thrift2004movement,karnath2001spatial}, while \textit{embodied cognition} emphasizes the role of physical and sensory experiences in cognitive processes \cite{clark1998being,borghi2010embodied,sanchez2016embodied,saini2022somaflatables}. These concepts have influenced technology design in education \cite{oliveira2011haptic,phutane2022tactile,melaku2016interlocking,teshima2010three,ghodke2019cross,capovilla2013teaching,ludi2014accessible}, socialization \cite{bhowmick2017insight,hornecker2006getting,nettleingham2018community,fisher2012tangible}, and daily activities \cite{giudice2018navigating,shi2020molder,rosenblum2015braille}.

As we have seen, a considerable amount of effort has been put on accessible technology for BLV people; however, accessible technologies for music practice is still overlooked for this population. Indeed, while in the last few years, there has been a growing interest in accessible music technology - a range of devices have been developed to accommodate users with physical \cite{larsen2016prospects} or mental disabilities \cite{ooms2024promoting} - little research focused on BLV people. In two key publications published in 2018 and 2019, Frid systematically analyzed accessible DMIs \cite{frid2018accessible,dd2019accessible} and reported that ``Little research in the community appears to have focused on developing musical interfaces specifically for persons who are blind'' \cite{dd2019accessible}.  In 2025, we found a number of new papers published after 2019 that propose technology aimed at enhancing music practice for BLV individuals. Given that music is a highly valuable activity for BLV persons, we argue that it is crucial to map out the current state of the discussion on accessible music technology for this community. 

\section{Methodology}
\label{sec:methdology}

To uncover music technology tailored to BLV people, we conducted a systematic literature review, following PRISMA method - a widely used reviewing method in HCI research \cite{hansson2021decade,scuri2022hitting}. 
The identification of our paper corpus includes three main steps: (1) identifying relevant databases, (2) conducting a keyword search, and (3) screening and excluding articles from the corpus. 
We analyzed the papers by progressively clustering them to identify types of study and technology. 
In figure \ref{fig:method}, we outline the process of how we followed the PRISMA and clustered the results.

\subsection{Defining the corpus of papers using PRISMA} 

\subsubsection{Identification of database.}
We selected the ACM Digital Library\footnote{\url{https://dl.acm.org/}} and IEEE repository \footnote{\url{https://ieeexplore.ieee.org/Xplore/home.jsp}} as these are the two main publishers for interactive computing research. Additionally, we included venues focused on music technology, such as the New Interfaces for Musical Expression (NIME) Conference Proceedings Archive \footnote{\url{https://nime.org/archives/}}, the Journal of New Music Research \footnote{\url{https://www.tandfonline.com/journals/nnmr20}} and the International Journal of Human-Computer Studies \footnote{\url{https://www.sciencedirect.com/journal/international-journal-of-human-computer-studies }}. 

\subsubsection{Keyword search.}
We defined a set of keywords to include papers that overlap music technology and BLV people. We used ``music'' as a keyword and did not include sound, as we specifically wanted to focus on music and not on aural feedback or auditory display. For the population we included ``blind'' OR ``visually impaired'' OR ``low vision'' OR ``visual impairments''. In ACM library, IEEE library and International Journal of Human-computer Studies we used both (``blind'' OR ``visually impaired'' OR ``low vision'' OR ``visual impairments'' AND ``music''). In New Interfaces for Musical Expression (NIME) and New Music Research, we used only the keyword related to BLV individuals (``blind'' OR ``visually impaired'' OR ``low vision'' OR ``visual impairments'') as these venues already focus on music. 
The search for key terms was independently performed on title, abstract, and author keywords. We used the search engine provided by the online repositories:
Zenodo \footnote{\url{https://zenodo.org/}}, 
ACM Digital Library \footnote{\url{https://dl.acm.org/}}, 
IEEE Xplore \footnote{\url{https://ieeexplore.ieee.org/Xplore/home.jsp}},
Taylor \& Francis Online \footnote{\url{https://www.tandfonline.com/}}, 
and Science Direct \footnote{\url{https://www.sciencedirect.com/}}
This initial search was conducted in March 2025 and produced an initial corpus of 477 results, after excluding duplication (e.g., the same paper coming up both when the search was conducted in abstract or in keyword), we identified 142 unique papers.

\subsubsection{Screening and exclusion criteria.}
We reviewed all abstracts for screening and applied the following exclusion criteria: 
\begin{itemize} 
\item Incorrect population: 49 excluded papers where the term blind referred to computational techniques. 
\item Not focused on music: 87 excluded papers that focus on sound rather than music, such as 11 papers using audio to represent visual content (data sonification) or 76 paper use audio as an auditory interface (sonic cues, earcons, auditory icons), without a focus on music. 
\item Paper format: 6 excluded student doctoral symposium papers or short pieces of writing proposing future research. 
\end{itemize}

After applying these exclusion criteria to the initial set of 142 papers, the number of relevant articles was narrowed down to 54 unique entries, focusing specifically on technology design for the BLV people.

\begin{figure*}[ht]
\centering
\includegraphics[width=0.9\textwidth]{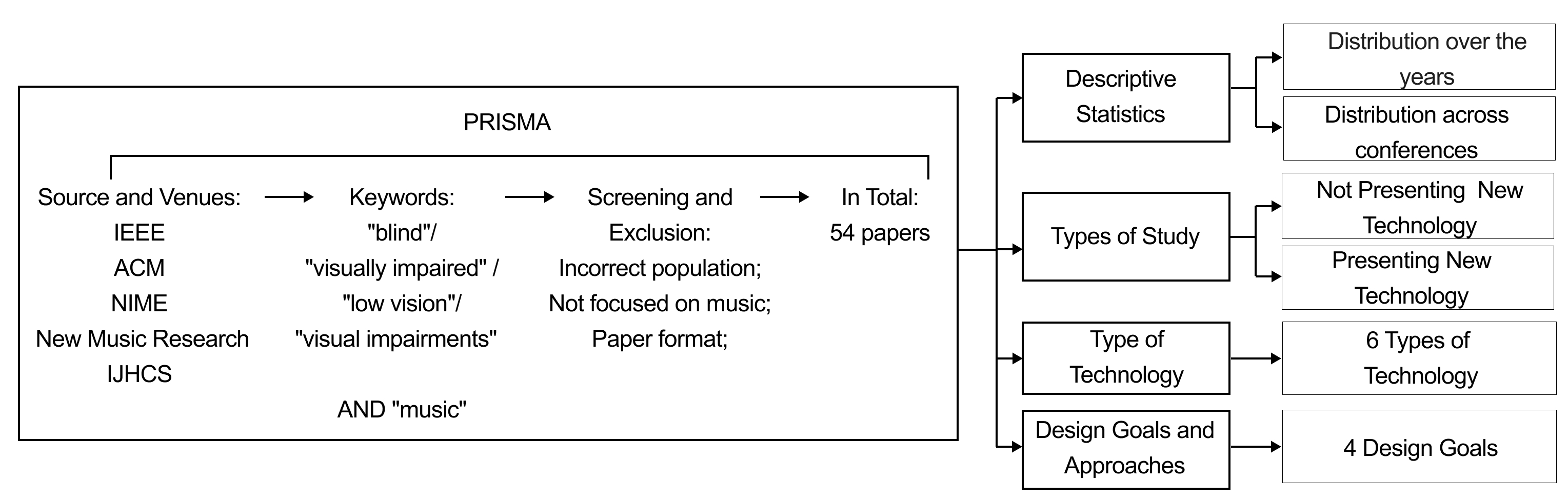}
\caption{Overview of the methods to collect the paper following the PRISMA guidelines and of the clustering process of systematic literature review. We also highlight the results emerging from the different methods.}
\label{fig:method}
\Description{This figure shows the process of building a literature review. We followed the PRISMA guidelines for a systematic review of 54 papers. Sources include IEEE, ACM, New Interfaces for Musical Expression (NIME), New Music Research and International Journal of Human-computer Studies. Keywords like "blind," "low vision," and "music" were used. Screening excluded irrelevant populations or formats. Finally, we categorized papers into descriptive statistics, types of study, six types of technology, and design goals and approaches.}
\end{figure*}

\subsection{Analysis of the corpus of papers}

We addressed our general research goal by following a structured set of steps (overview of paper collection and analysis in figure \ref{fig:method}). To initially familiarize ourselves with the corpus, we extracted the metadata of the 54 papers into a spreadsheet. The spreadsheet included details such as titles, authors, publication venues, years, research methods, research contributions, and a brief description of the technology developed. We will present descriptive statistics to showcase the distribution of papers by year within this corpus.

After running the descriptive statistics outlining publication trends, we analyzed the papers using two criteria 1) \textit{type of studies} - considering the level of participation of BLV people in the study -; and 2) \textit{type of technology} - for which type of musical activity the technology was intended.

To identify the \textit{types of study}, analyze the type of studies, involvement of BLV individuals, and number of participants, we used the same approach proposed in a recent meta analysis of user engagement in CHI (which we use as a benchmark) \cite{caine2016local}, thus marking for all the paper number of participants, study type (for this we deductively used the list of different study types proposed by \cite{caine2016local}), study length. Given the highlighted importance of focusing on social inclusivity for accessible technology (see background), in addition to that paper we also looked at the study length and study environment (where there was some engagement in real-world context). We also checked the overall frame of the study (where it was evaluating a new technology, or investigating existing issues or aspects) and in case of new technology at what stage BLV people were involved (only evaluation or in a co-design process plus evaluation). For each paper we copied the respective type of study, environment, and number of participants, lengths of studies in a table (see \ref{tab:studytype}).

To identify the \textit{type of technologies} proposed, and the respective trends, we recursively coded the papers using a methodology for clustered literature derived by thematic analysis \cite{onwuegbuzie2016mapping}, similar to other recent literature reviews \cite{bergstrom2021evaluate}. We initially coded each of the selected papers separately, using open-ended text summarizing the dimensions we were looking at - as done in a recent literature review presented at CHI \cite{bergstrom2021evaluate}. The codes were then recursively harmonized and clustered till identifying the categories we will present later. To ensure that this was done consistently, the first author initially coded all the papers, and another author double checked the process. While clustering the codes we constantly rechecked the original papers in case of doubts. The clustering has been re-discussed till reaching agreement.

\section{Results}
\label{sec:Res}

We present the results of three analyses conducted: (1) descriptive statistics of paper distribution across years and publication venues; (2) \textit{type of studies} - 
taking into consideration the level of engagement of BLV people in the study -; and 3) \textit{type of technology} - for what type of music activity the technology was designed for.

\subsection{Descriptive statistics}
\begin{figure*}[ht]
\centering
\includegraphics[width=0.9\textwidth]{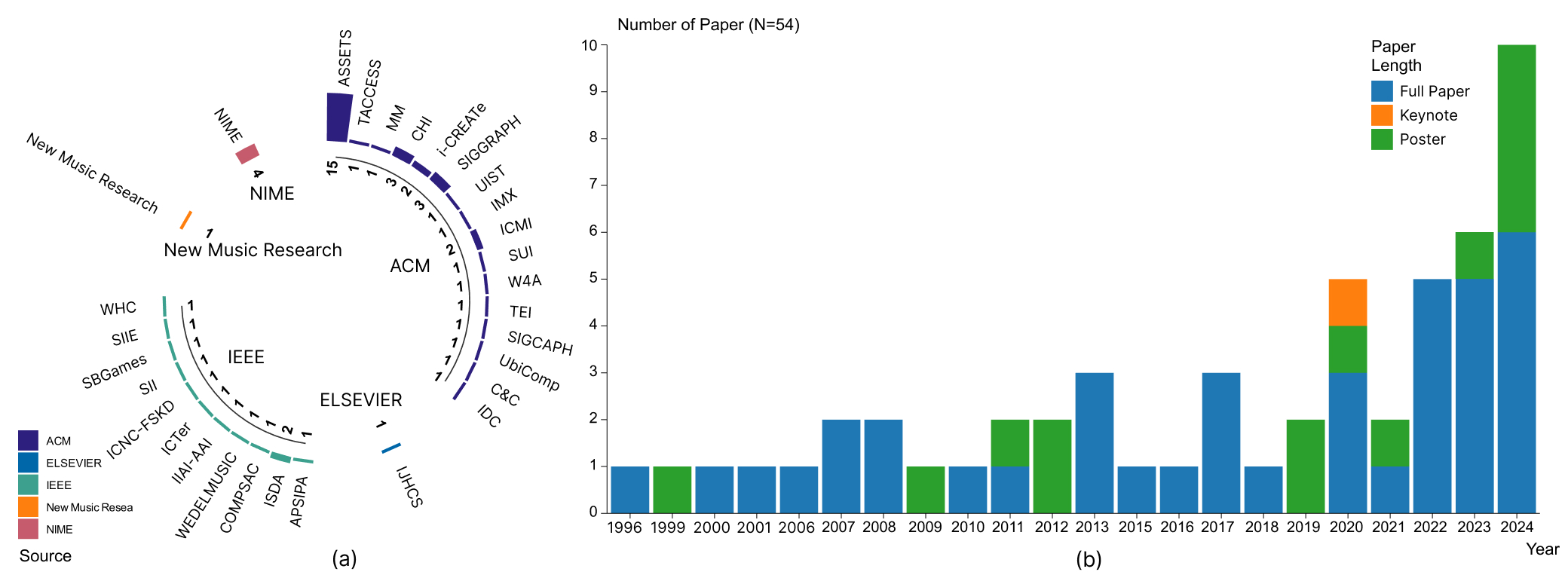}
\caption{(a) Distribution of papers across different conferences; (b) Distribution of papers across different years.}
\label{fig:circle}
\Description{(a) Distribution of the 54 papers we reviewed across different conferences. ASSETS conference stands out because of 15 papers about musical technology. (b) Distribution of types of paper and its published year. The figure shows an increase in the number of papers related to musical technology for BLV people after 2019.}
\end{figure*}

In Figures \ref{fig:circle}, we provide an overview of the paper's distributions. 
As shown in Figure \ref{fig:circle}, the majority of papers have been presented at the ASSETS conference, which is unsurprising given its prominence as the primary venue for research on accessible technology.

As seen in Figure \ref{fig:circle}, the last 10 years have shown a significant increase in studies, with 28 papers (almost 51.85\% of the total) published after the recent literature review on assistive music technology by Frid in 2019 \cite{dd2019accessible}. In that review, Frid pointed out that music technology for BLV people remains under-investigated. While this claim still holds true, it is important to emphasize that research focused on BLV individuals has become more prolific in recent years, indicating growing interest and advancements in this area.

\subsection{Types of Paper and Study}

We describe here the different types of paper and types of study in this subsection. Types of paper are divided by the main topics of the paper, but one paper might have more than one study. To discriminate the paper and study, we follow the existing study to count the exact user studies \cite{caine2016local}. First we present the papers that did not involve the development of new technology, and then we present the paper with new technological advancement. Finally, we provide an overview of the methodologies used regardless of technological development. (All papers details with paper types title, reference and short summary see \ref{tab:appendix})

\subsubsection{\textbf{Papers not presenting new technologies.}}
Ten papers (N=10) investigated the engagement of BLV individuals with music technology, without designing new devices. We categorized these studies into three main groups: 1) papers that conducted studies to gather insights for the design of future devices (N=6); 2) studies focusing on existing devices for making music (N=3); and 3) a keynote (N=1). 

The six papers that conducted studies to gather design insights for future devices explored diverse directions. Volta et al. studied motion synchronization differences between BLV and non-BLV musicians \cite{volta2021sensorimotor}; Lu et al. investigated the use of vibrotactile patterns and materials to assist BLV people in making music \cite{lu2023playing} ; Lu et al. explored different multimodal interactions to improve readability of music scores \cite{lu2024we}; Dang et al. collected data to explore the opportunity of creating VR musical concerts for BLV people \cite{dang2023opportunities,dang2024towards}; and Park et al. examined differences in emotional perception between BLV and non-BLV individuals, proposing design solutions based on their findings \cite{park2020comparative}.

Among the three studies focusing on existing devices for making music, two studies interviewed BLV individuals to examine how they use available technology \cite{payne2020blind,lu2023there} . In contrast, Payne et al. tested an accessible piece of music technology in a real-world setting \cite{payne2022empowering}. This technology, developed by the authors, had been previously presented \cite{payne2022soundcells}.

Finally, one keynote paper reviewed the advancements in music technology for BLV individuals over the past few years \cite{tanaka2020musical}.

\begin{table*}[ht]
\centering
\caption{Overview of the types of study. Paper marked with * presented work on collaboration between BLV and non-BLV people.}
\Description{This table summarizes two types of study related to musical technology, including studies presenting new technology and studies not presenting new technology. It lists the total papers for each type, whether they involve no participants, professional BLV musicians, non-musician BLV participants, non-BLV participants, or a mix. It also notes if studies were tested in real-world contexts, with citations.}
\label{tab:studytype}

\small
\setlength{\tabcolsep}{3pt}
\sloppy
\begin{tabular}{
|>{\raggedright\arraybackslash}m{1.5cm}
|>{\raggedright\arraybackslash}m{2.2cm}
|>{\raggedright\arraybackslash}m{1cm}
|>{\raggedright\arraybackslash}m{1.4cm}
|>{\raggedright\arraybackslash}m{1.4cm}
|>{\raggedright\arraybackslash}m{1.4cm}
|>{\raggedright\arraybackslash}m{1.4cm}
|>{\raggedright\arraybackslash}m{1.4cm}
|>{\raggedright\arraybackslash}m{1.4cm}|}
\hline
\multicolumn{2}{|l|}{} & & & \multicolumn{4}{c|}{Different Levels of Participants’ Involvement} & \\ \cline{5-8}
\multicolumn{2}{|l|}
{\multirow{-2}{*}{Types of Study}} & 
\multirow{-2}{*}{\begin{tabular}[c]{@{}c@{}}Total\\Number\end{tabular}} & 
\multirow{-2}{*}{\begin{tabular}[c]{@{}c@{}}No\\Participant\end{tabular}} & 
\begin{tabular}[c]{@{}c@{}}Professional\\BLV\\Musician\end{tabular} & 
\begin{tabular}[c]{@{}c@{}}BLV not\\Characte-\\rized by\\Their Music \\ Skills\end{tabular} & 
\begin{tabular}[c]{@{}c@{}}Mixed BLV\\and\\non-BLV\end{tabular} & 
\begin{tabular}[c]{@{}c@{}}Only\\non-BLV \\ People\end{tabular} & 
\begin{tabular}[c]{@{}c@{}}Tested in \\ Real-World \\ Context\\ (a subset \\of previous \\ categories)\end{tabular} \\ \hline

\multirow{3}{*}{\begin{tabular}[c]{@{}l@{}}Presenting\\New\\Technology\end{tabular}} 
& Design Only & 13 & 
\begin{minipage}[t]{\linewidth}\raggedright
12 papers \cite{10.1145/1328491.1328539}, \cite{abe2008braille}, \cite{lackner1999sensory},\\
\cite{mannone2019cubeharmonic}, \cite{kanai2011pocopoco}, \cite{malliopoulos2001music},\\
\cite{kobayashi2022music}, \cite{homenda2010intelligent}, \cite{luckner2006braille},\\
\cite{housley2013implementation}, \cite{huang2018research}, \cite{xie2024acoustic}
\end{minipage} & 
/ & 
/ & 
1 paper \cite{dang2024musical} & 
/ & 
/ \\ \cline{2-9}

&\multirow{1}{*}{\begin{tabular}[c]{@{}l@{}} Design and \\ Evaluate\end{tabular}} & 23 & 
/ & 
\begin{minipage}[t]{\linewidth}\raggedright
3 papers \cite{payne2019non}, \cite{manenti2024accessibility}*, \cite{payne2024different}*
\end{minipage} & 
\begin{minipage}[t]{\linewidth}\raggedright
9 papers \cite{anderson1996composability}, \cite{kim2011tapbeats}, \cite{costa2022framework},\\
\cite{ranasinghe2017non}, \cite{de2015virtual}, \cite{vetter2020welle},\\
\cite{payne2023live}*, \cite{payne2023approaches}*, \cite{liu2024noteblock}
\end{minipage} &
\begin{minipage}[t]{\linewidth}\raggedright
7 papers \cite{haenselmann2009tangible}, \cite{omori2013collaborative}*, \cite{capozzi2012musica},\\
\cite{yuan2008blind}, \cite{miller2007finger}, \cite{challis2000weasel}, \cite{nishida2022digitusync}
\end{minipage} &
\begin{minipage}[t]{\linewidth}\raggedright
4 papers \cite{kawarazaki2014supporting}, \cite{yoo2017longitudinal}, \cite{ueda2024tactile},\\
\cite{dimogerontakis2024musicane}*
\end{minipage} & 
\begin{minipage}[t]{\linewidth}\raggedright
4 papers \cite{haenselmann2009tangible}, \cite{payne2023live}*, \cite{payne2023approaches}*, \cite{payne2024different}*
\end{minipage} \\ \cline{2-9}

& \begin{tabular}[c]{@{}l@{}}Co-Design or\\Formative Study,\\and Evaluate\end{tabular} & 8 & 
/ & 
\begin{minipage}[t]{\linewidth}\raggedright
2 papers \cite{turchet2021musical}*, \cite{payne2022soundcells}
\end{minipage} & 
1 paper \cite{tanaka2016haptic} & 
\begin{minipage}[t]{\linewidth}\raggedright
3 papers \cite{10.1145/3173574.3173643}, \cite{sabuncuoglu2020tangible}, \cite{yairi2012music}*
\end{minipage} & 
2 papers \cite{turchet2023give}, \cite{ding2024redesigning} & 
1 paper\cite{tanaka2016haptic} \\ \hline

\multirow{3}{*}{\begin{tabular}[c]{@{}l@{}}Not\\ Presenting\\New\\Technology\end{tabular}} 
& \begin{tabular}[c]{@{}l@{}}Collecting Insights\\for Future Design\end{tabular} & 6 & 
/ & 
2 papers \cite{volta2021sensorimotor}*, \cite{lu2024we} & 
3 papers \cite{lu2023playing}, \cite{dang2024towards}, \cite{dang2023opportunities} & 
1 paper \cite{park2020comparative} & 
/ & 
/ \\ \cline{2-9}

& Studying Existing Devices & 3 & 
/ & 
3 papers \cite{payne2020blind}, \cite{lu2023there}, \cite{payne2022empowering} & 
/ & 
/ & 
/ & 
1 paper \cite{payne2022empowering} \\ \cline{2-9}

& Keynote & 1 & 
1 paper \cite{tanaka2020musical} & 
/ & 
/ & 
/ & 
/ & 
/ \\ \hline
\end{tabular}
\end{table*}

\subsubsection{\textbf{Papers presenting new technology.}}
As shown in Table \ref{tab:studytype}, 44 papers described projects where a new technological system was developed. We clustered each paper based on the level of participants’ involvement. The subsequent sections classify the papers from the least involvement (papers focused solely on technical development) to the highest involvement (papers involved participants in total sessions of co-design or formative study, and evaluation). 

Among the papers (N=13) which designed new technologies \textbf{without involving participants in evaluation phase}, twelve papers (N=12) have no participant in the evaluation or development phases \cite{10.1145/1328491.1328539,abe2008braille,lackner1999sensory,mannone2019cubeharmonic,kanai2011pocopoco,malliopoulos2001music,kobayashi2022music,homenda2010intelligent,luckner2006braille,housley2013implementation,huang2018research,xie2024acoustic}. Most of these publications are posters or demos. Additionally, one paper (N=1) involved domain experts to consult design insights but still does not include participant in the evaluation.

The majority of papers (N=23) presented a new musical device \textbf{design and evaluation} but did not include co-design activities or formative activities. Twelve papers (N=12) involved only BLV participants, with three focusing on professional BLV musicians \cite{payne2019non,manenti2024accessibility,payne2024different}, and the remaining nine targets lay BLV individuals \cite{anderson1996composability,kim2011tapbeats,costa2022framework,ranasinghe2017non,de2015virtual,vetter2020welle,payne2023live,payne2023approaches,liu2024noteblock}. Seven other papers (N=7) involved both BLV and non-BLV participants \cite{haenselmann2009tangible,omori2013collaborative,capozzi2012musica,yuan2008blind,miller2007finger,challis2000weasel,nishida2022digitusync} for different reasons: one of these studies \cite{omori2013collaborative} was designed for collaboration between the two population, four studies involved non-BLV people in public events or to generalize the technology or the results \cite{haenselmann2009tangible,capozzi2012musica,yuan2008blind,nishida2022digitusync}, finally one study \cite{challis2000weasel} recruited non-BLV participants because it was hard to find blind musicians.

The remaining three studies (N=4) evaluated systems solely with sighted participants, \cite{kawarazaki2014supporting,yoo2017longitudinal,ueda2024tactile,dimogerontakis2024musicane}. In detail, \cite{kawarazaki2014supporting,dimogerontakis2024musicane} tested their system with blindfolded non-BLV people; \cite{ueda2024tactile} tested a method to convey conductor's instruction to blind musicians, but did not involve visually impaired people; \cite{yoo2017longitudinal} tested their system with non-BLV participants and suggest that their system would work well for visually impaired people too; \cite{ding2024redesigning} involved the non-BLV teachers from schools for BLV people.

Finally, eight papers (N=8) involved BLV participants in \textbf{a co-design or formative study and} evaluated the prototype systems with participants. Five of these studies \cite{payne2022soundcells,turchet2023give,yairi2012music,sabuncuoglu2020tangible,10.1145/3173574.3173643} followed a formative process, including design and evaluation: one study recruited only professional BLV musicians \cite{payne2022soundcells}; another study recruited non-BLV participants in formative study to gather general design insights; the other three included both BLV and non-BLV participants who played different roles in the design \cite{yairi2012music,sabuncuoglu2020tangible,10.1145/3173574.3173643}. The remaining three studies \cite{tanaka2016haptic,turchet2021musical,ding2024redesigning} implemented three workshops as part of an iterative design process, with one targeting BLV people with professional musical skills \cite{turchet2021musical} and the other focusing on lay BLV people \cite{tanaka2016haptic}. For all the papers involving BLV people in the studies we also checked if the technology was tested in real-world contexts and if there any collaborative musical activities - an overview of the study types presented in table \ref{tab:studytype}.

\subsubsection{\textbf{From different type of paper to individual studies.}}

\begin{figure*}[ht]
\centering
\includegraphics[width=\linewidth]{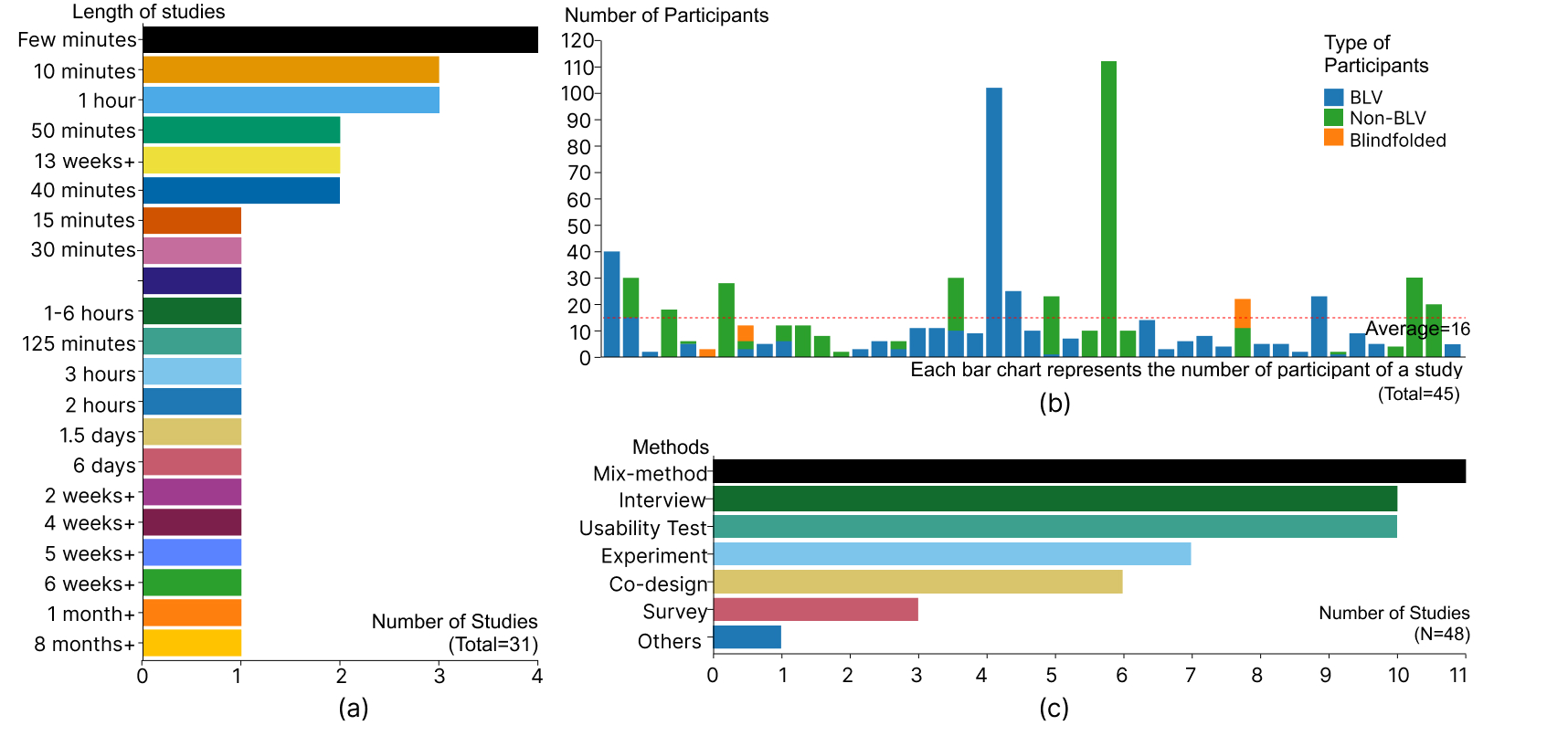}
\caption{The number of participants, methods used, and study length were reported in the papers we reviewed. (a) shows the distribution by the number of participants reported (N=45). (b) shows the distribution by the number of methods used (N=48). (c) shows the distribution by the study length (N=31) reported.}
\label{fig:allbar}
\Description{The number of participants, methods used, and study length were reported in the papers we reviewed. (a) shows the distribution by the number of participants reported, and 45 papers report the exact number of participants. (b) shows the distribution by the number of methods used, and 48 papers report the methods used. (c) shows the distribution by the study length, while 31 papers report the exact duration.}
\end{figure*}

To compare the papers we reviewed with the standards of general HCI research mentioned above, we examined the types of studies in detail, regardless of paper type. In general HCI research, the existing standard for distinguishing papers and studies is that a study can use mixed methods, but it must involve the same group of participants \cite{caine2016local}. Otherwise, when different groups of participants are involved, the studies within a paper are considered separate studies using various methods. Thus, we counted the number of individual studies (N=63) following the existing standard. We reviewed all the existing studies to understand how studies involving human participants were conducted. Studies’ details of each paper (methods used, number of participants, study length of each study) are visualized in Figure \ref{fig:allbar}.

Among 45 studies involving participants, 48 reported the study methods used. Some papers (N=4) only gave a vague description of the informal trial sessions \cite{payne2019non,ranasinghe2017non,kim2011tapbeats, anderson1996composability}. 
On average, studies recruited 16 participants (Mean=16), see Figure \ref{fig:allbar}, a value that aligns closely with the mean reported by Caine for HCI studies \cite{caine2016local}. Among the reported methodologies, mixed methods and usability testing emerged as the two most frequently employed approaches. Usability testing was used to evaluate whether specific functionalities met their intended purposes (see Figure \ref{fig:allbar}). Notably, these tests did not differ significantly from standard usability tests in HCI research, except that their primary focus was related to vision impairments. Mixed methods, as defined by Caine, involve combining two or more methods with the same group of participants \cite{caine2016local}. It is particularly noteworthy that researchers working with music for BLV people often felt the need to combine multiple methods to gather insights from participants, being higher than the typical distribution in CHI research, where mixed-method is used in only 5\% of studies \cite{caine2016local}. While no individual method stood out as exceptional, the fact that many researchers decided to combine more than one method underscores the value of integrating multiple methods when working with BLV participants to obtain more comprehensive data.
Study durations were generally brief or at least constraints to one session (see Figure \ref{fig:allbar}); of the 31 studies that reported activity lengths, the majority (N=21) described sessions lasting from a few minutes to a few hours. A few studies (N=4) reported their duration of not specific but short time, there are two situations for studies under 10 minutes: first, some studies involved musical performances or compositions reported as short; second, experimental trials where we calculated the session length to be just a few minutes. Eight papers detailed longer-term engagements, involving co-design activities and evaluations of existing technologies and prototypes over several weeks \cite{tanaka2016haptic,payne2022soundcells,turchet2021musical,payne2022empowering,manenti2024accessibility,payne2023live,payne2023approaches,payne2024different}. Additionally, six studies reported conducting their research in real-world settings \cite{haenselmann2009tangible,tanaka2016haptic,payne2022empowering,payne2023approaches,payne2023live,payne2024different}. This represents a major limitation to the state of the art, which we will discuss in a dedicated section \ref{sec:socio}.

\subsection{Types of Technology} 

\begin{figure*}[ht]
\centering
\includegraphics[width=\textwidth]{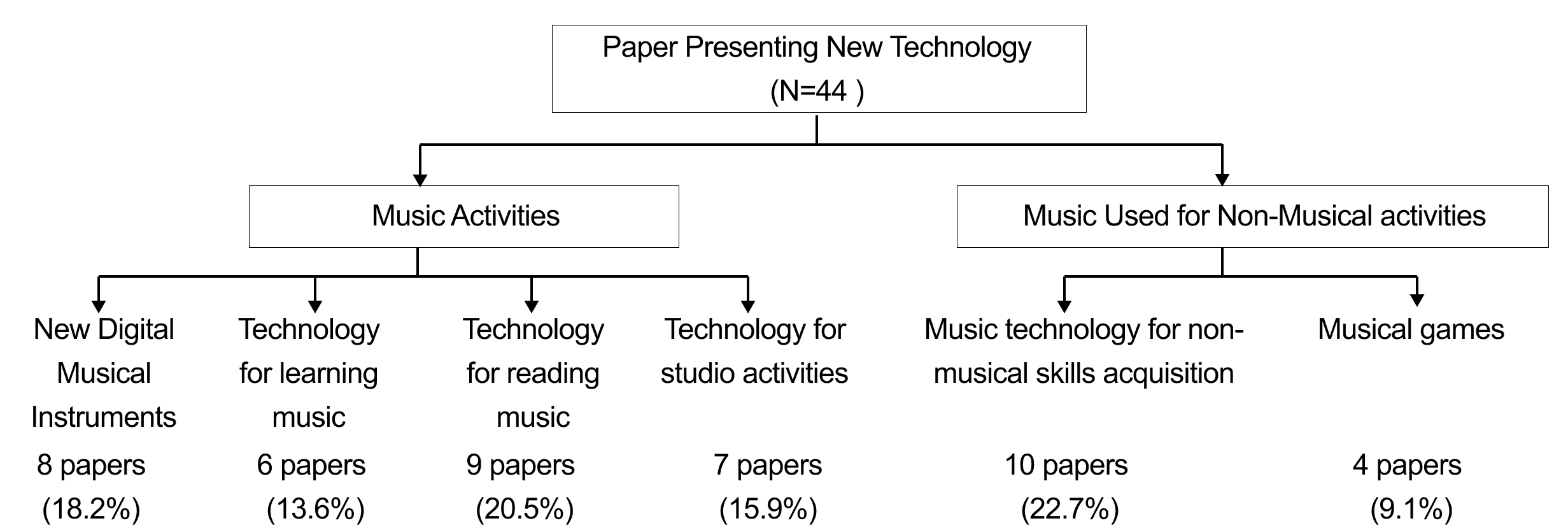}
\caption{Types of technology we cluster by music activities, four categories - New Digital Musical Instruments, Technology for Learning Music, Technology for reading music, and technology for studio activities - are designed for musical activities, while the remaining two categories -music technology for non-musical skill acquisition and musical games - rely on music used for non-musical activities.}
\label{fig:diagram}
\Description{Types of technology we cluster by music activities, four categories - New Digital Musical Instruments, Technology for Learning Music, Technology for reading music, and technology for studio activities - are designed for musical activities, while the remaining two categories -music technology for non-musical skill acquisition and musical games - rely on music used for non-musical activities.}
\end{figure*}

In this section, we provide an overview of all the technological artifacts presented in the corpus of the research papers we identified, clustered by type of technology (see Figure \ref{fig:diagram}). Table \ref{tab:techtype} provides an overview of the types of technological artifact clustered by target activity and highlights the population involved in the related studies.

\begin{table*}[ht]
\centering
\caption{Summary of the types of technology. Paper marked with * presented work on collaboration between BLV and non-BLV people.}
\Description{This table summarizes different musical technologies and how they have been evaluated in research. It shows the total papers for each technology, whether they were tested without participants, with target participants, co-designed, tested with non-BLV participants only, or tested in real-world settings. Additionally, two papers marked with asterisks present work on collaboration between BLV and non-BLV people. Citations are provided for each category.}
\label{tab:techtype}

\small
\setlength{\tabcolsep}{3pt}
\sloppy
\begin{tabular}{
|>{\raggedright\arraybackslash}m{2.2cm}
|>{\raggedright\arraybackslash}m{1cm}
|>{\raggedright\arraybackslash}m{2cm}
|>{\raggedright\arraybackslash}m{2cm}
|>{\raggedright\arraybackslash}m{2cm}
|>{\raggedright\arraybackslash}m{2cm}
|>{\raggedright\arraybackslash}m{2cm}|}
\hline
& & & \multicolumn{3}{c|}{\textbf{Evaluate with Participants}} & \\ \cline{4-6}
\multirow{-2}{*}{\textbf{\begin{tabular}[c]{@{}c@{}}New Musical \\ Technology\end{tabular}}} & 
\multirow{-2}{*}{\textbf{\begin{tabular}[c]{@{}c@{}}Total \\ Number\end{tabular}}} & 
\textbf{\begin{tabular}[c]{@{}c@{}}No Participant\\Involved in \\Evaluation\end{tabular}} & 
\textbf{\begin{tabular}[c]{@{}c@{}}Tested with\\Target \\ Participants\end{tabular}} & 
\textbf{\begin{tabular}[c]{@{}c@{}}Co-design\\and\\Evaluate\end{tabular}} & 
\textbf{\begin{tabular}[c]{@{}c@{}}Tested with\\Only non-BLV\\People\end{tabular}} & 
\textbf{\begin{tabular}[c]{@{}c@{}}Tested in the\\Real-world \\Context\\(subset of \\ left columns)\end{tabular}} \\ \hline

\begin{tabular}[c]{@{}l@{}}New Digital\\Musical\\ Instruments\end{tabular} & 8 & 
\begin{minipage}[t]{\linewidth}\raggedright
3 papers \cite{kanai2011pocopoco}, \cite{kobayashi2022music}, \cite{xie2024acoustic}
\end{minipage} & 
\begin{minipage}[t]{\linewidth}\raggedright
3 papers \cite{haenselmann2009tangible}, \cite{omori2013collaborative}*, \cite{payne2019non}
\end{minipage} & 
1 paper \cite{yairi2012music}* & 
1 paper \cite{dimogerontakis2024musicane}* & 
1 paper \cite{haenselmann2009tangible} \\ \hline

\begin{tabular}[c]{@{}l@{}}Technology for\\Learning Music\end{tabular} & 6 & 
/ & 
\begin{minipage}[t]{\linewidth}\raggedright
2 papers \cite{nishida2022digitusync}, \cite{liu2024noteblock}
\end{minipage} & 
1 paper \cite{turchet2021musical} & 
\begin{minipage}[t]{\linewidth}\raggedright
3 papers \cite{kawarazaki2014supporting}, \cite{yoo2017longitudinal}, \cite{ueda2024tactile}
\end{minipage} & 
/ \\ \hline

\begin{tabular}[c]{@{}l@{}}Technology for\\Reading Music\end{tabular} & 9 & 
\begin{minipage}[t]{\linewidth}\raggedright
6 papers \cite{10.1145/1328491.1328539}, \cite{abe2008braille}, \cite{homenda2010intelligent},
\cite{luckner2006braille}, \cite{housley2013implementation}, \cite{huang2018research}
\end{minipage} & 
\begin{minipage}[t]{\linewidth}\raggedright
2 papers \cite{challis2000weasel}, \cite{manenti2024accessibility}*
\end{minipage} & 
1 paper \cite{payne2022soundcells} & 
/ & 
/ \\ \hline

\begin{tabular}[c]{@{}l@{}}Technology for\\Studio Activities\end{tabular} & 7 & 
1 paper \cite{malliopoulos2001music} & 
\begin{minipage}[t]{\linewidth}\raggedright
4 papers \cite{capozzi2012musica}, \cite{anderson1996composability},\\ \cite{ranasinghe2017non}, \cite{vetter2020welle}
\end{minipage} & 
1 paper \cite{tanaka2016haptic} & 
1 paper \cite{turchet2023give} & 
1 paper \cite{tanaka2016haptic} \\ \hline

\begin{tabular}[c]{@{}l@{}}Music Technology \\ for Non-musical\\Skills Acquisition\end{tabular} & 10 & 
2 papers \cite{lackner1999sensory}, \cite{mannone2019cubeharmonic} & 
\begin{minipage}[t]{\linewidth}\raggedright
7 papers \cite{milne2017blocks4all}, \cite{costa2022framework}, \cite{payne2023live}*, \cite{payne2023approaches}*,\\ \cite{payne2024different}*, \cite{ding2024redesigning}, \cite{dang2024musical}
\end{minipage} & 
1 paper \cite{sabuncuoglu2020tangible} & 
/ & 
3 papers \cite{payne2023live}*, \cite{payne2023approaches}*, \cite{payne2024different}* \\ \hline

Musical Games & 4 & 
/ & 
\begin{minipage}[t]{\linewidth}\raggedright
4 papers \cite{yuan2008blind}, \cite{miller2007finger}, \cite{kim2011tapbeats}, \cite{de2015virtual}
\end{minipage} & 
/ & 
/ & 
/ \\ \hline
\end{tabular}
\end{table*}

\subsubsection{\textbf{Technology for reading music.}}
The first category we clustered incorporates those papers (N=9) that present technology for reading music notation. Three papers described systems that can transform different music notation formats, such as MusicXML or MIDI input, into Braille music scores \cite{luckner2006braille,abe2008braille,10.1145/1328491.1328539}. Homenda extended this possibility by relying on AI transcription to transfer also printed music notation (as well as MIDI files, and Music XML files) to Braille scores \cite{homenda2010intelligent}. Huang et al. used a CNN model in their system to recognize Braille music \cite{huang2018research} . Payne et al. relied on braille in a browser-based music notation software that incorporates text-to-speech feedback and features for screen reader compatibility \cite{payne2022soundcells}. The last two papers that deal with notation did not rely on Braille but implemented a multimodal approach. Housley et al. developed a mobile application that integrates a friendly layout, color, and interaction for low vision people to access sheet music on tablets relying on their residual vision \cite{housley2013implementation}. Additionally,  Challis and Edwards presented a multi-modal system that combines a tactile interface and audio output designed to access music notation \cite{challis2000weasel}. Finally, Manenti and Ardan presented 3D pinted tactile graphic scores for BLV people to read music and collaborate \cite{manenti2024accessibility}.

Many papers highlighted that the learning curve for reading music notation is notably steep with both standard Braille and specialized music Braille scores \cite{huang2018research,homenda2010intelligent,abe2008braille,10.1145/1328491.1328539,payne2022soundcells}. Additionally, complications increase with multiple melodic lines \cite{abe2008braille}. Therefore, it is important to investigate other forms of music representation. Applying this approach directly to music reading can be useful. As reading can be relevant also to other activities, we will dedicate one design insight (subsection \ref{sec:access_scores}) to notation within the context of accessing information.

\subsubsection{\textbf{Technology for learning music.}} 
The second category we identified encompasses those papers (N=6) that presented technology for learning traditional music performance skills. Three papers (N=3) focused on synchronization, mainly for orchestral or chorus practice and performance, enhancing non-verbal communication between conductor and performer. kawarazaki wt al. developed an electronic music baton that can detect conductors’ movements and provide the movements’ information to BLV singers via haptic vibrations, tested only with non-BLV conductors \cite{kawarazaki2014supporting}. Similarly, Ueda et al. presented that vibrotactile actuators have been used to convey the conductor's hand and baton movements to musicians, tested with non-BLV participants \cite{ueda2024tactile}. The remaining paper probed collaboration among BLV individuals via a series of studies to improve synchronization through vibrotactile armbands, belts, and bracelets that provide feedback on tempo, volume and timing \cite{turchet2021musical}. The remaining two papers (N=2) focused on individual practice. For instance, an augmented glove has been designed to aid teachers in conveying correct hand positions by adjusting to the user's hand movements \cite{nishida2022digitusync}. Additionally, Liu et al. proposed a tactile 3D interface for learning music notes and improving music literacy \cite{liu2024noteblock}. Finally, vibrotactile actuators on the user's arms have been used to provide tuning feedback through variations in vibration intensity, reducing the need for visual cues, tested with non-BLV participants \cite{yoo2017longitudinal}.

\subsubsection{\textbf{Technology for studio activities.}}
Seven papers (N=7) presented technology for studio activities and focused on composing, recording, and editing exploring two main solutions, tangible interfaces \cite{vetter2020welle,tanaka2016haptic,capozzi2012musica,anderson1996composability} or text-to-speech feedback \cite{ranasinghe2017non,malliopoulos2001music}. Concerning tangible interfaces, Anderson presented a system for composing and editing music with simplified mouse clicks and keyboard shortcuts to facilitate access for BLV users \cite{anderson1996composability}. Similarly, Capozzi and colleagues presented an application for producing music, that implemented editing shortcuts directly on a keyboard \cite{capozzi2012musica}. Finally, Tanaka et al. developed a haptic digital audio workstation in a wooden board that employed motorized sliders and an infrared sensor for audio editing via tactile interaction \cite{tanaka2016haptic} - this is the only prototype tested in a real-world context as producers' home studio. While this study proposed technology for studio context, the tactile displaying also improve the readability of music scores.

Moving on to text-based interaction, Ranasinghe and Jayaratne developed a synthesized singing model that facilitates composing and editing via non-visual text-based cues \cite{ranasinghe2017non}. Vetter developed a web-based platform that creates and records music through a simplified text-based interface controlled with a computer keyboard \cite{vetter2020welle}. Additionally, Malliopoulos et al. designed a system that relies on text-to-speech feedback to facilitate music editing, while also mimicking other software environments to minimize the learning curve \cite{malliopoulos2001music}. Finally, Turchet et al. proposed an online platform to search, play and create music \cite{turchet2023give}.

All these strategies proved to be good for helping BLV people to effectively act in the studio environment. However, as pointed out in one of the paper thatnot presenting technologies there is an overall problem to keep the assistive devices up-to-date due to rapid advancements in professional studio equipment and applications, which tend to result in obsolescence of the accessible interfaces\cite{payne2020blind}. There is a need to do more work to cope with this tendency, in the papers we analyzed this issue is still open and deserves further investigation.

\subsubsection{\textbf{New digital musical instruments (DMI)}}
The fourth category we identified consists of eight papers (N=8) that presented new Digital Musical Instruments (DMIs) designed for BLV people. Two papers (N=2) introduced new controllers: one developed a tactile board for playing loop-based music, similar to a MIDI sequencer \cite{kanai2011pocopoco}; the other presented a set of blocks with different musical functions (start, play notes and rest) as USB music controllers \cite{kobayashi2022music}. Two additional papers (N=2) relied on shortcuts: one designed a piano keyboard equipped with shortcuts to change tonalities \cite{haenselmann2009tangible}, and the other developed a web-based drum sequencer controlled by shortcuts on a computer keyboard \cite{payne2019non}. The remaining two DMIs (N=2) were specifically intended for collaborative purposes. One study introduced three cards used as controllers, where their movements were mapped to piano, guitar, and percussion sounds \cite{yairi2012music} . The other study presented a device that allows two individuals to create music by manipulating cards connected by ropes (mapped to seven basic music loops) and a dice (mapped to six different musical instruments) \cite{omori2013collaborative}. These six studies collectively showcase a diverse range of approaches to designing DMIs for BLV individuals, highlighting innovations in tactile interfaces, however, more work is needed to actually build musical practice around these tools. The last two papers (N=2) focused on mapping the spatial structure to the music sequencer \cite{dimogerontakis2024musicane,xie2024acoustic}: Dimogerontakis et al. facilitated the spatial issues with cane, the most familiar tool for BLV people \cite{dimogerontakis2024musicane}; Xie et al. proposed a  musical instrument which used the real world scenes \cite{xie2024acoustic}.

\subsubsection{\textbf{Music technology for non-musical skills acquisition.}}
The fifth category we identified includes papers (N=10) that introduced new tangible educational instruments aimed at fostering creative music composition and organization activities to enhance abstract thinking, creativity, and programming skills. Two papers(N=2) focused on abstract thinking and social communication. Lackner et al. developed a sensory puzzle to provide auditory and tactile feedback during a music creation experience, promoting both creative thinking and social communication through the reorganization of musical patterns \cite{lackner1999sensory}. Another study proposed a Rubik’s cube whose faces were mapped to musical notes and chords, enabling users to generate music as a way to stimulate abstract thinking \cite{mannone2019cubeharmonic}.

The remaining three papers focused on leveraging musical tangible interfaces to teach programming skills(N=3). For instance, Costa et al. introduced tangible blocks that can be rearranged to create melodic sequences, supporting the development of computational thinking \cite{costa2022framework}. Similarly, Sabuncuoglu presented tangible blocks of varying shapes for composing melodies, aimed at facilitating the understanding of programming concepts and computational thinking \cite{sabuncuoglu2020tangible}. Finally, Milne and Ladner designed tangible blocks embedded with music to teach programming languages, such as Scratch and Blockly, through block reorganization \cite{10.1145/3173574.3173643}.
Four papers (N=4) engaged coding into music editing. Live coding alongside music performance is way to understand and enjoy music while gaining computational skills. Payne et al. presented a long-term series of studies in real context at school for cultivating live coding by collaborative music performing \cite{payne2024different,payne2023approaches,payne2023live}. Additionally, Ding et al. co-designed with teachers of BLV people a browser-based application for content browsing, music and code editing \cite{ding2024redesigning}.

Finally, one paper proposed the musical appreciation and learning in virtually reality with AI audio description of instrument and environment settings to help with a immersive musical performance \cite{dang2024musical}.

\subsubsection{\textbf{Musical games.}}
Finally, the sixth category includes those papers (N=4) that presented musical video games for BLV individuals. For instance, a 3D audio environment allows players to have a better immersive experience while playing musical instruments in the game by developed binaural audio techniques \cite{de2015virtual}. Another paper proposed a game based on musical rhythms for Android \cite{kim2011tapbeats}. Similarly, Miller and colleagues presented an original audio-based rhythm-action game designed to entertain BLV individuals and facilitate collaboration with non-BLV people \cite{miller2007finger}. The last paper introduced a haptic glove that provides an alternative approach to playing the existing musical games with tactile stimuli \cite{yuan2008blind}. 

The results of studies on music video games highlight the importance of social engagement and inclusivity \cite{de2015virtual,kim2011tapbeats,miller2007finger,yuan2008blind}, with BLV users feeling included in mainstream gaming. Overall, learned from these papers, future video game designs should prioritize these aspects to foster social inclusion, ensuring accessibility and engagement for all users and enriching the experience across communities.
% Atkinson et al. emphasize that designing games with accessibility in mind enhances enjoyment and benefits broader interpersonal interactions \cite{atkinson2006making}. 

% The results of all the papers about music video games consistently highlight the importance of social engagement and inclusivity \cite{de2015virtual,kim2011tapbeats,miller2007finger,yuan2008blind}, and BLV users report feeling included when participating in mainstream gaming. Atkinson et al. point out that designing games with accessibility in mind not only benefits game enjoyment but also extends to broader interpersonal domains \cite{atkinson2006making}. Thus, future efforts in video game design should actively integrate these aspects to foster true social inclusion, ensuring accessibility and engagement for all users, as well as enriching the gaming experience across different communities. 

\subsection{Design Goals and Approaches}
\label{sec:goals}

Our literature review has uncovered trends in how this research has tackled musical experiences for BLV people. Here, we present four design goals, challenges, and approaches derived from the papers we analyzed.

\subsubsection{Fostering spatial awareness in musical instruments playing.} BLV people compensate for missing visual cues through touch and memory, but sometimes they experience cognitive overload. To minimize cognitive load, designs reduce simultaneous modalities \cite{sabuncuoglu2020tangible} and employ tangible anchors (e.g.\ foot-pedals or linked objects) to stabilize hand placement \cite{haenselmann2009tangible,yairi2012music}. Building on familiar layouts such as keyboards further eases orientation \cite{haenselmann2009tangible,payne2019non}. These principles extend beyond digital musical instruments to studio gear, games, and learning tools.

\subsubsection{Accessing Music Content}
\label{sec:access_scores}
Accessing visual information about musical content further complicates performance. The learning curve for both standard and music-Braille notation is steep \cite{huang2018research,homenda2010intelligent,abe2008braille,10.1145/1328491.1328539,payne2022soundcells,payne2022empowering}. Additional challenges include simultaneous melodic lines \cite{abe2008braille} and overly sparse scripts \cite{challis2000weasel}. Recent assistive-technology studies emphasise the need for intuitive information access, pointing toward fast-learning, simplified scores or alternative representations that bypass Braille. Abstract, intuitive shapes on physical objects aid BLV interaction \cite{sabuncuoglu2020tangible,manenti2024accessibility}, while touchable sound-wave displays effectively convey music in studio settings \cite{tanaka2016haptic}.

% The challenge of accessing visual information about musical content further complicates performance. The learning curve for reading music notation is notably steep with both standard Braille and specialized music Braille scores \cite{huang2018research,homenda2010intelligent,abe2008braille,10.1145/1328491.1328539,payne2022soundcells,payne2022empowering}. Additionally, some other issues specifically related to music notation do exist (e.g., multiple melodic lines simultaneously \cite{abe2008braille}, too sparse scripts \cite{challis2000weasel}).
%  We derive a design goal from recent studies on assistive technologies that designs need to support intuitive access to information. While existing studies explore simpler methods of accessing musical content, a clear future direction emerges—there is a need for fast learning and simplified musical scores or notations. Alternative forms of music representation that entirely bypass Braille also appear promising. 
%  Research has demonstrated that using abstract and intuitive primitive shapes for physical objects can facilitate interaction for BLV users \cite{sabuncuoglu2020tangible,manenti2024accessibility}. Additionally, Tanaka and Parkisson highlighted how touchable representations of soundwaves can effectively represent music in a studio context \cite{tanaka2016haptic}. 

\subsubsection{Facilitating (Non-verbal) Communication in Collaborative Performance}

Researchers have explored auditory \cite{lackner1999sensory} and vibrotactile haptic feedback \cite{ueda2024tactile,yoo2017longitudinal,turchet2021musical}, with haptic feedback showing promising results \cite{anderson1996composability,tanaka2016haptic,yuan2008blind}. However, the quality of materials (supporting breathability, malleability, and durability) is crucial when designing wearable haptic devices for BLV people to ensure a positive experience \cite{lu2023playing}. While these solutions have mainly been explored in music education, they are essential for synchronization in musical performance, making them relevant for designing new musical instruments. Additionally, non-verbal communication strategies can inspire collective musical games, though the high cognitive load during play must be considered.

% Music performers in orchestras or groups often rely on non-verbal communication \cite{volta2021sensorimotor}, such as visual cues and eye contact \cite{fredrickson1994band}, which can be challenging for BLV players who struggle to receive these signals from each other \cite{lu2023there,baker2016perceptions} and from conductors \cite{turchet2021musical}. To address this issue, researchers have explored auditory \cite{lackner1999sensory} and vibrotactile haptic feedback \cite{ueda2024tactile,yoo2017longitudinal,turchet2021musical} as potential solutions, with haptic feedback showing particularly promising results \cite{anderson1996composability,tanaka2016haptic,yuan2008blind}. It is however important to underline that the quality of the materials (supporting breathability, malleability, and durability) plays an important role in designing wearable haptic devices for BLV people \cite{lu2023playing}, to create a good experience for BLV people. Although these solutions have primarily been explored in music education, synchronization is vital for playing any instrument, making these strategies relevant when designing new musical instruments. Moreover, non-verbal communication strategies could also inspire collective musical games. However, also in this case it is necessary to account for the high level of cognitive load focus while playing.

\subsubsection{Supporting Memory}
Supporting memory is also a crucial goal for BLV people to engage effectively with music, as they cannot read music while playing \cite{park2020comparative}. Recent literature underscores the importance of supporting memory in musical interfaces for BLV people, using tactile and vibrational feedback to enhance musical engagement. Some of the papers we analyzed have highlighted how tactile braille and vibrational feedback facilitate the conversion of musical notation into muscle memory, significantly enhancing recall capabilities \cite{lu2023there,yuan2008blind,omori2013collaborative}.

\section{Design Insights}
\label{sec:discussion}

For each approach we propose several design insights combining the systematic literature review with aspects of assistive technology for BLVs presented in the background (subsection \ref{bk_tecBLV}). Table \ref{tab:goalftech} gives an overview and  highlights how each of these approaches relates to the different types of technology. Therefore, we discuss \textbf{D}esign \textbf{I}nsights (DIs) based the research and technologies facilitate the current design goals and approaches.

\begin{table}[ht]
\centering
\caption{An examination of how various technological approaches can contribute to four design goals, where ``X'' represents applicability of the technology to the respective design goal.}
\Description{This table presents an examination of how various technological approaches can contribute to four design goals. The technologies for DMIs can help with the design goals of fostering spatial awareness, facilitating (non-verbal) communication and supporting memory. The technologies for learning music can also be an effective approach to foster spatial awareness, access music content, and facilitate (non-verbal) communication. Technologies for reading music can facilitate better experience in accessing music content and supporting memory. Technologies for studio activities can help foster spatial awareness, access music content, and support memory. Technologies for non-music skills can be an effective approach to foster spatial awareness and access music content. Last, technologies for musical games can foster spatial awareness.}
\label{tab:goalftech}
\resizebox{\linewidth}{!}{%
\begin{tabular}{|l|l|l|l|l|} 
\hline
Technology\textbackslash{}Goal & \begin{tabular}[c]{@{}l@{}}Fostering \\Spatial Awareness\end{tabular} & \begin{tabular}[c]{@{}l@{}}Accessing \\Music Content\end{tabular} & \begin{tabular}[c]{@{}l@{}}Facilitating (Non-verbal) \\Communication\end{tabular} & \begin{tabular}[c]{@{}l@{}}Supporting \\Memory\end{tabular} \\ 
\hline
Technologies for DMIs & ~ ~ ~ ~ ~ ~ ~ X &  & ~ ~ ~ ~ ~ ~ ~ ~ ~ X & ~ ~ ~ ~ ~ ~ ~ X \\ 
\hline
Technologies for~Learning Music & ~ ~ ~ ~ ~ ~ ~~X & ~ ~ ~ ~ ~ X & ~ ~ ~ ~ ~ ~ ~ ~ ~~X &  \\ 
\hline
Technologies for~Reading Music &  & ~ ~ ~ ~ ~~X &  & ~ ~ ~ ~ ~ ~ ~~X \\ 
\hline
Technologies for~Studio Activities & ~ ~ ~ ~ ~ ~ ~~X & ~ ~ ~ ~ ~~X &  & ~ ~ ~ ~ ~ ~ ~~X \\ 
\hline
Technologies for~Non-music Skills & ~ ~ ~ ~ ~ ~ ~~X & ~ ~ ~ ~ ~~X &  &  \\ 
\hline
Technologies for~Musical Games & ~ ~ ~ ~ ~ ~ ~~X &  &  &  \\
\hline
\end{tabular}
}
\end{table} 

\subsection{DI1: Fostering Spatial Awareness by Reducing Cognitive Overloads} 

Spatial awareness underpins effective interaction with any interface \cite{thrift2004movement,karnath2001spatial}. Tactile components in BLV-focused instruments serve a dual role: they orient players and support the complex motor actions of performance \cite{altenmuller2006music,magnusson2009epistemic}. Spatial awareness remains a core hurdle across assistive-technology research \cite{oliveira2011haptic,phutane2022tactile,shi2020molder,teshima2010three,ghodke2019cross,bhowmick2017insight,dimogerontakis2024musicane,zhang2024beadwork,el2013touch,baldwin2017tangible}, where tactile input consistently improves navigation \cite{berla1977tactual,hasper2015methods,capovilla2013teaching,milne2018blocks4all,ludi2014accessible,melaku2016interlocking,rosenblum2015braille,aldrich2001tactile}. Since unfamiliar digital music instruments can overwhelm cognition loads \cite{pigeon2019cognitive,loomis2018sensory}, solutions from the New Interfaces for Musical Expression (NIME) community may help \cite{fmorreale2017,nime2023_14}. Tactile materials, vital in BLV people's daily lives, offer a promising avenue for designing music technology for them. To be specific, designers should therefore link physical elements—e.g., straps, foot pedals, or rigid couplings—to create continuous tactile maps \cite{haenselmann2009tangible,yairi2012music}, and reuse familiar embodied layouts such as piano keys to keep cognitive load low \cite{haenselmann2009tangible,payne2019non,branje2014playing}.

\begin{figure*}[tbh]
\centering
\includegraphics[width=\textwidth]{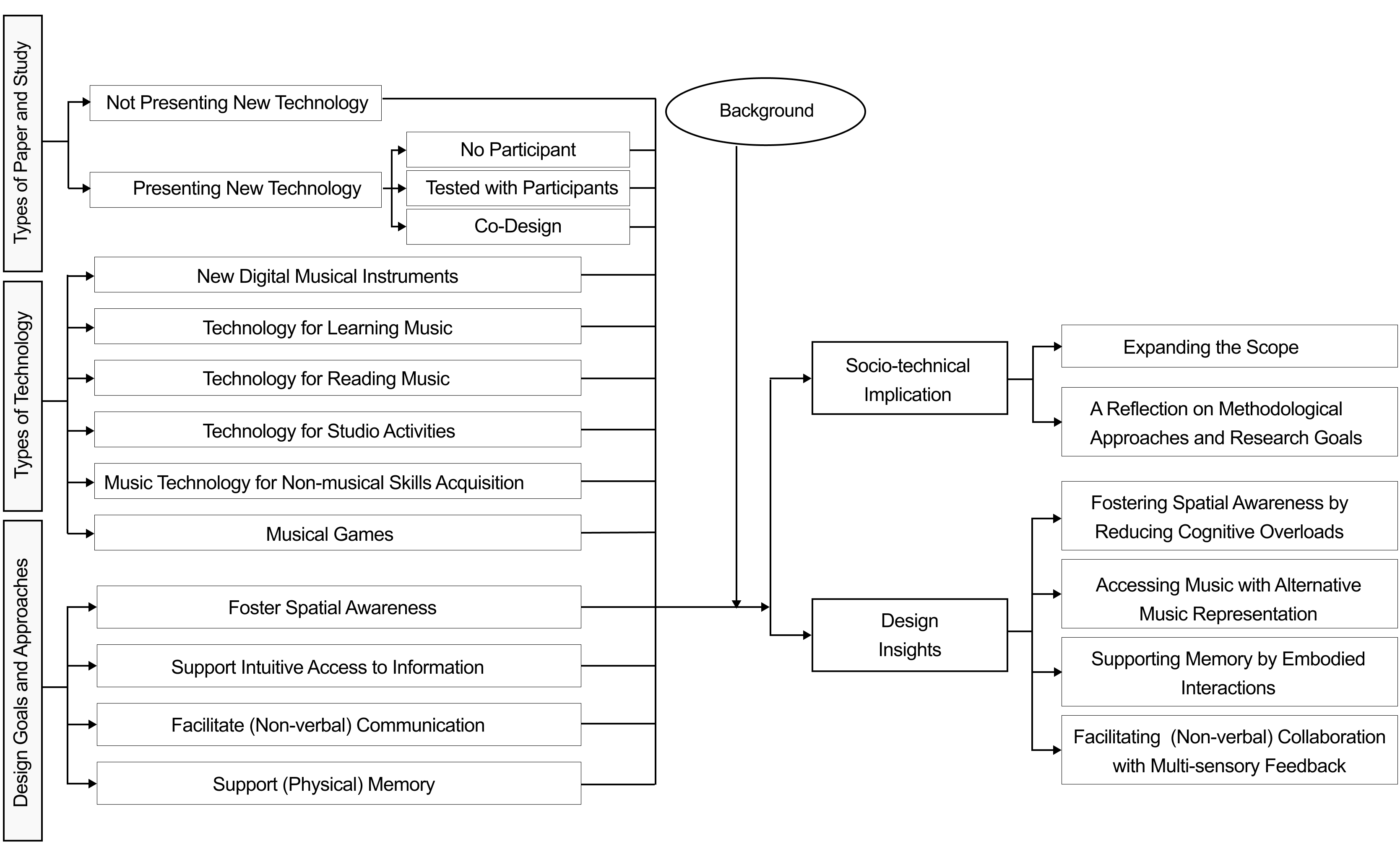}
\caption{An overview of the connections between the different types of technology and the four approaches.}
\label{fig:dis}
\Description{An overview of the connections between the results and the discussion. Based on the results that present the types of study and technology, and design goals and approaches, we propose four key insights into designing accessible music technology for BLV people. Finally, we discuss the need to shift from merely addressing accessibility challenges to actively promoting social inclusivity.}
\end{figure*}

\subsection{DI2: Accessing Music with Alternative Music Representation}

\begin{figure*}[ht]
\centering
\includegraphics[width=0.9\textwidth]{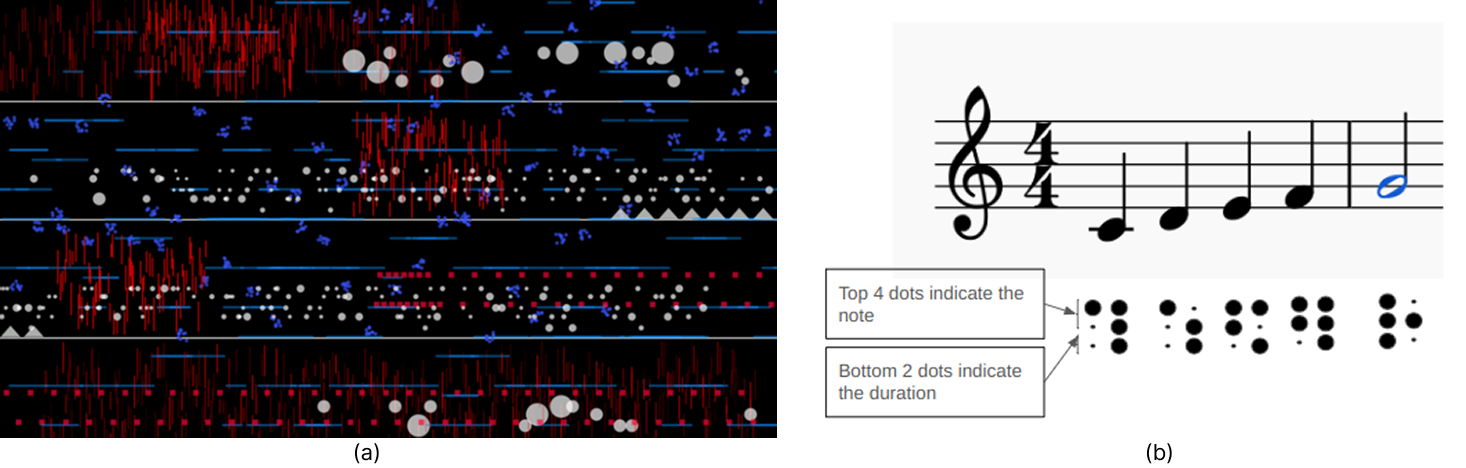}
\caption{(a) A simple music scale in standard music notation (on top), and with Braille notation (bottom). (b) A new form of score was invented for electronic music representation \cite{dal2022exploring}. Permission to reproduce it obtained by the authors.}
\label{fig:scoreBrraille}
\Description{(a) presents an example of a simple music scale in standard music notation (on top), and with Braille notation (bottom). (b) shows a new form of score was invented for electronic music representation. Permission to reproduce it obtained by the authors.}
\end{figure*}

Performing music is a complex activity that demands a high level of focus, coordination, and cognitive effort \cite{lu2023there,loomis2018sensory}. The importance of intuitive design is even more relevant when it comes to music representation and notation. Indeed, accessing music representation is of uttermost importance for acquiring music skills. Braille music relies on the standard six-position Braille cell, invented by Louis Braille, but assigns different meanings, related to music position, to the various combinations of the six dots.  This method is widely used, and big repositories of Braille music exist (e.g., Braille scores in the International Music Score Library Project (IMSLP)\footnote{\url{https://imslp.org/wiki/Category:Works_with_Braille_scores}} or Braille music collection at the Library of Congress in the United States\footnote{see \url{https://www.loc.gov/nls/services-and-resources/music-service-and-materials/ }}.). However, the learning curve to master this form of notation is very steep, and can be detrimental for amateur or entry-level BLV people (figure \ref{fig:scoreBrraille} showcases a simple scale with Braille). Learning braille is difficult in everyday life and educational contexts \cite{abramo2013ethnographic,abramo2013ethnographic,jessel2015access,baker2016perceptions}.

Exploring alternatives to traditional notation can mitigate these barriers. Park argues that Braille scores require faster-learning substitutes to reach wider audiences \cite{park2015useful}. Abstract scores in prior studies \cite{sabuncuoglu2020tangible,manenti2024accessibility,tanaka2016haptic} and avant-garde practices offer guidance. Cage’s Notation compiled diverse experimental formats \cite{cage2012silence}, and subsequent work has greatly expanded the field \cite{masu2021nime}. For instance, figure \ref{fig:scoreBrraille}, from a recent study on electronic-music performance, depicts music with simple shapes to give musicians an intuitive grasp of notes \cite{dal2022exploring}.

% Tackling alternatives to traditional musical scores and notation can be useful to overcome Braille music scores' limitation. Park have emphasized that Braille music scores need alternatives to support fast learning and broader spreading \cite{park2015useful}. Similar to the abstract music scores used in the existing papers \cite{ sabuncuoglu2020tangible,manenti2024accessibility,tanaka2016haptic}, contemporary avant-garde music can provide inspiration for alternative notation systems and suggest alternatives. Contemporary literature on musical scores has highlighted how a vast form of devices and styles of representation can be used as a score.  Cage collected an extensive set of examples in his book Notation \cite{cage2012silence} in the late sixties. Since then the work on alternative notations has greatly expanded \cite{masu2021nime}. As an example, figure \ref{fig:scoreBrraille} is taken from a recent study on alternative music representations for performing electronic music. In that work music is represented using simple forms aiming to provide musicians with an intuitive representation of the music notes \cite{dal2022exploring}.

\subsection{DI3: Supporting Memory by Embodied Interactions}

Music performers in orchestras or groups often rely on non-verbal communication, such as visual cues and eye contact \cite{volta2021sensorimotor}, which can be challenging for BLV players who struggle to perceive these signals \cite{lu2023there,baker2016perceptions} and from conductors \cite{turchet2021musical}. Engaging embodied interaction in music technology development is an effective approach to support memorizing music scores while playing. Embodied cognition, which suggests that cognitive processes are deeply rooted in the body's interactions with the world \cite{clark1998being,borghi2010embodied,sanchez2016embodied}, further supports the use of embodied interaction in learning and memorizing \cite{saini2022somaflatables,medina2017epistemic,tomavs2014concept,magnusson2009epistemic}. Since BLV musicians cannot rely on visual cues to read music while playing, tactile braille and vibrational feedback have been effective in translating musical notations directly into muscle memory, a connection that is pivotal for enhancing recall during performances \cite{lu2023there,park2020comparative,yuan2008blind}. The connection between muscle memory and music practice is well-documented. For instance, Lam pointed out that ``music production and muscle movement are so interconnected that the process of creating music must consider the physicality behind auditory perceptions \cite{lam2020physicality}.''

Overall, we propose this insight on prioritizing embodied interaction to improve the readability of music scores and enrich the musical experience for BLV individuals. Additionally, this insight can mutually benefit DI 1 and DI 2.

% Combining the literature about muscle memory in music with the insights on the BLV people's embodied cognition, we support that relying on tangible and embodied experience supporting the acquisition of muscle memory is crucial for BLV people's music practices. 

% Thus, integrating embodied experiences into musical practice not only aligns with how BLV people process information but also enhances the physicality that is inherent in music production, as noted by Lam \cite{lam2020physicality}, eventually leading to ease the mnemonic processes. Consequently, leveraging tangible and embodied learning methods is essential for supporting BLV individuals in their musical endeavors.

\subsection{DI4: Facilitating  (Non-verbal) Collaboration with Multi-sensory Feedback}
Multi-sensory feedback systems can bolster collaborative performance for BLV musicians. Evidence shows that engaging multiple senses benefits collaboration \cite{bhowmick2017insight}. Such systems should adapt to performers’ needs and conditions, integrating vibratory and auditory cues \cite{ueda2024tactile,yoo2017longitudinal,turchet2021musical,lackner1999sensory,anderson1996composability,tanaka2016haptic,yuan2008blind}.

Wearable devices—armbands or belts with actuators—embed tactile cues directly into musicians’ attire without sacrificing style \cite{yoo2017longitudinal,turchet2021musical,lu2024we,lu2023playing}. Real-time analytics using sensors for tempo, pitch, and other parameters enable immediate adjustments and cohesive ensemble output \cite{borchers2004personal,welch2005real}. Collaborative feedback loops that let musicians exchange cues further enhance group synchronisation. Consequently, multi-sensory systems for BLV musicians should prioritise adaptability and real-time interaction.

Most existing prototypes target teacher–student coordination \cite{ueda2024tactile,kawarazaki2014supporting,turchet2021musical}. Although haptic feedback proves effective, these systems have not been trialled in music schools. Testing in authentic learning environments is essential to validate educational impact \cite{harmes2016framework}; we identify this as a key avenue for future work.

\section{What is Missing? Toward a Socio-technical Implication from Music Technology for BLV People}
\label{sec:socio}

In this subsection, we contextualize the music technology tailored to BLV people within the recent switch from conceiving disability as an individual medical issue to understanding it as a social construct \cite{hoppestad2007inadequacies,mack2021we,siebers2008disability}. 
With this reflection, we do not want to undermine the importance of accessibility, which remains a key component of inclusivity.

\subsection{Expanding the Scope: From Music Activities for Social Connections}
Here we propose four possible points that would better integrate existing music technologies, approaches and goals within social contexts of BLV people’s lives.  Considering the demonstrated role of music in supporting programming education, it is plausible that these disciplines could also benefit from integrating music technology. Lastly, it is worth noting that a few papers involved studies conducted in a real school environment\cite{milne2017blocks4all, payne2024different, payne2023approaches,payne2023live}.
As previously discussed, engaging with real-world learning contexts is essential, suggesting that future research should prioritize such practical applications\cite{harmes2016framework}.

First, it is necessary to look into ways to integrate BLV people more effectively, mitigating the obsolescence of research projects, into the current music industry by leveraging advancing technologies. 
The possibility of using tools for editing and composing is also very important as it allows BLV people to express themselves by creating new music or even working in the music industry \cite{lu2023there,payne2020blind}. 
However, as pointed out by Payne due to the rapid updates in professional studio equipment and applications, assistive devices designed for accessibility tend to become obsolete quickly \cite{payne2020blind}. 
These issues arguably cannot be effectively addressed through design alone, but need to be considered within the socioeconomic context of software updates. 
Recent literature on sustainable research products has highlighted how open source software can effectively contrast obsolescence, and can probably inspire new solutions \cite{masu2024sustainable,masu2023nime,bettega2022off}. 

Second, BLV people can naturally build social connections by collaborative playing \cite{omori2013collaborative,yairi2012music}. 
Tangibility has already been proven to be good for social connection \cite{hornecker2006getting}. 
The various approaches to support making music that we identified - creating new musical instruments and technology to support learning,  reading, composing, or recording - can all contribute to developing social connections while making music. 
However, these potentials are rarely studied. 
Therefore, there is a need to further explore it in practice. 
Payne and colleagues conducted few studies where they actually engaged with a school \cite{payne2022soundcells,payne2024different, payne2023approaches,payne2023live}. 
More research in this direction is needed.

Third, we have seen that music can play a significant role in promoting social inclusivity within non-musical educational contexts \cite{lackner1999sensory,mannone2019cubeharmonic,costa2022framework,sabuncuoglu2020tangible,10.1145/3173574.3173643}. 
Studies in education showed that music can be fruitful in promoting social inclusivity \cite{crawford2020socially,baker2016perceptions}.
Indeed, innovative tools that utilize music-based tactile interfaces seem promising in enhancing creative and abstract thinking \cite{costa2022framework,mannone2019cubeharmonic,lackner1999sensory,yuan2008blind}, as well as in developing programming skills \cite{sabuncuoglu2020tangible,10.1145/3173574.3173643}, thus eventually helping BLV people find a place in society. 
However, also in this case, research tended to stop at the prototyping phase, and no study focusing on using music to develop non-musical skills was actually conducted in a real-world educational setting.

Finally, music-based video games can serve as powerful platforms for fostering social connections. 
Games that emphasize musical elements over visual content or tasks facilitate BLV individuals to engage more easily in them \cite{kim2011tapbeats,de2015virtual}. 
These games are particularly useful thanks to their simplicity and social engagement \cite{yuan2008blind,miller2007finger}. 
However, this inclusivity is typically a byproduct rather than an intentional consideration in the game design process. 
All these goals need to be considered in the design process to ensure social connection in inclusivity and accessibility, making it necessary to discuss their methodological implications. 
We elaborate on this in the next subsection.

\subsection{A Reflection on Methodological Approaches and the Research Aims}

Our results highlight that most papers focused on designing new devices (34 out of 42). 
Of these 34, 11 studies did not involve any evaluation or testing, and 3 papers did not include any participation of BLV individuals. 
Previous research in psychology and perception has noted that relying on blindfolded and sighted participants is invalid due to differences in physical perception between BLV, blindfolded, and sighted individuals \cite{postma2007differences,heller1989texture}. In total, 14 out of the 34 new technologies presented were not tested with BLV participants, and only three papers lasted more than one section or engaged in real-world contexts \cite{tanaka2016haptic,payne2022soundcells,haenselmann2009tangible}. One of these \cite{payne2022empowering} tested a device previously developed by the same authors \cite{payne2022soundcells}. While this specific case is commendable for its deep engagement with the population, the general trend is problematic. Current work on assistive technology for BLV people increasingly stresses that longitudinal testing is vital for understanding both personal development and social impact \cite{de2025sensing,adams2016blind,mathis2025lifeinsight,rector2017design}.
For example, Silva and colleagues evaluated their prototype across an extended series of dance classes to promote sustained engagement, exploration, and learning \cite{de2025sensing}.
More broadly, HCI research recommends longitudinal approaches for evaluation and long-term user engagement \cite{minton2025longitudinal,wang2025facilitating,ambe2022collaborative}.

Recently Rodgers and Marshall pointed out that examining technology usage in real-world settings is necessary to truly understand how people integrate these tools into their lives \cite{rogers2017research}. In light of the aforementioned challenges with technology adoption in real social contexts, it is even more important to foster the adoption of technology in everyday life rather than merely focusing on building new devices \cite {baker2023disability}. As Shinohara emphasized, it is not sufficient to simply create state-of-the-art technology to meet the assistive needs of BLV individuals \cite{shinohara2017design}. According to the authors, it is indeed essential to account for people's self-esteem and public perception when designing new technologies and to consider how these technologies will fit into their overall lives \cite{shinohara2017design}. Increasing reliance on technology without considering social factors may ultimately be counterproductive \cite{Foley2012technology, li2021choose, profita2016effect}. Indeed, individuals with disabilities often exhibit low acceptance and high abandonment rates of assistive technologies \cite{hurst2011empowering}. This issue has also been discussed in the context of music technology \cite{masu2023nime,morreale2017design}, and more broadly in relation to technology adoption \cite{hurst2011empowering,Foley2012technology, li2021choose,10.1145/3537797.3537875,10.1145/3536169.3537787,bettega2021s}. The critique of a technocentric approach argues that merely focusing on technological advancement, rather than on its actual societal usage, can be detrimental \cite{alexander2019critique,krier1985easy}. Introducing an excessive amount of technology requires significant investment, which, from a macro-social perspective, may not always be the most worthwhile solution \cite{clark2016technology}.

In practice, we suggest that studies that explore longer engagement with BLV people’s communities are needed. Recent literature relying on community studies \cite{bodker2016farmer}, or technometodology \cite{avila2019encumbered} published in HCI could be useful in this sense. For example, Bodker et al. engaged with a volunteer-based farmer community \cite{bodker2016farmer}, and Bettega et al. worked with local communities of environmentalists \cite{bettega2021s,capaccioli2016participatory}. In both cases, the studies relied on prolonged and engaged contacts with the communities. This approach could arguably be integrated into the design of music technology for BLVs to effectively consider how these technologies would fit into their overall lives \cite{shinohara2017design}. Another inspiration can be found in the studies engaging with grassroots activities. For instance, Teli and colleagues engaged with grassroots communities to understand needs and design for sustainable appropriation of technologies related to Radio \cite{teli2021understanding,teli2020tales}. Accounting for grassroots activities could be particularly relevant to genuinely engage with BLV people desires and hopes. 

\section{Conclusion}
This paper presents the first systematic review of literature on music technology for BLV individuals, classifying current studies by technology type and BLV involvement. We identify six main categories of music technology: 1) Technology for Reading Music, 2) Technology for Learning Traditional Music, 3) Technology for Studio Activities, 4) New Digital Musical Instruments, 5) Music Technology for Non-Musical Skills Acquisition, and 6) Music Technology and Video Games. We highlight a lack of studies engaging BLV people in real-world scenarios and discuss four key design goals: 1) foster spatial awareness, 2) support intuitive access to information, 3) facilitate (non-verbal) communication, and 4) support (physical) memory. These goals provide a foundation for accessible music technology design for BLV people.

The limitation of this review lies in the small sample size, but the increase in publications since 2019 suggests more opportunities for exploring the impact of state-of-the-art technologies on music accessibility. In conclusion, we argue that more work is needed to move from accessibility to inclusivity in music technology design for BLV people, with a focus on long-term studies in real-world contexts. Inclusivity, beyond accessibility, emphasizes social justice and participation, and future research should explore how music integrates into the broader lives of BLV individuals. Music should be viewed not only as a skill but as an activity that facilitates social integration and work opportunities.

\bibliographystyle{ACM-Reference-Format}
\bibliography{0-Main}

\appendix 
\section{Details of Surveyed Papers (Landscape)}

\clearpage
\onecolumn
\begin{landscape}
\small
\setlength{\tabcolsep}{4pt}
\sloppy
\begin{longtable}{|
    >{\raggedright\arraybackslash}m{2cm}|
    >{\raggedright\arraybackslash}m{2.5cm}|
    >{\raggedright\arraybackslash}m{4.5cm}|
    >{\centering\arraybackslash}m{1.3cm}|
    >{\raggedright\arraybackslash}m{10cm}|
}
\caption{Appendix: Overview of papers discussed in this Paper.}
\label{tab:appendix}\\
\hline
\textbf{Paper Types} & \textbf{Sub-types} & \textbf{Paper Title} & \textbf{Reference} & \textbf{Short Summary} \\ \hline
\endfirsthead
\multicolumn{5}{c}{{\bfseries \tablename\ \thetable{} -- continued}}\\\hline
\textbf{Paper Types} & \textbf{Sub-types} & \textbf{Paper Title} & \textbf{Reference} & \textbf{Short Summary} \\ \hline
\endhead
\hline \multicolumn{5}{r}{{Continued on next page}}\\\hline
\endfoot
\hline
\endlastfoot

% ——— Presenting New Technologies (44 行) ———
\multirow{44}{=}{Presenting New Technologies}

  % Technology for Reading Music (9)
  & \multirow{9}{=}{Technology for Reading Music}
    & Braille music score management environment
    & \cite{abe2008braille}
    & This paper described systems that can transform music notation formats from MusicXML into Braille music scores. \\ \cline{3-5}
  &  
    & MusicXML to Braille Music translation
    & \cite{10.1145/1328491.1328539}
    & This paper described systems that can transform different music notation formats, such as MusicXML and MIDI input, into Braille music scores. \\ \cline{3-5}
  &  
    & Braille Score
    & \cite{luckner2006braille}
    & The paper presents a developing computer program that helps the blind people dealing with music notation and MIDI file. \\ \cline{3-5}
  &  
    & Intelligent computing technologies in music processing for blind people
    & \cite{homenda2010intelligent}
    & The authors extended this possibility by relying on AI transcription to transfer also printed music notation (as well as MIDI files, and Music XML files) to Braille scores. \\ \cline{3-5}
  &  
    & Research on Braille Music Recognition Based on Convolutional Neural Network
    & \cite{huang2018research}
    & This paper used a CNN model in their system to recognize Braille music \\ \cline{3-5}
  &  
    & SoundCells: designing a browser-based music technology for braille and print notation.
    & \cite{payne2022soundcells}
    & This paper relied on braille in a browser-based music notation software that incorporates text-to-speech feedback and features for screen reader compatibility. \\ \cline{3-5}
  &  
    & Implementation Considerations in Enabling Visually Impaired Musicians to Read Sheet Music Using a Tablet
    & \cite{housley2013implementation}
    & Housley et al. developed a mobile application that integrates a friendly layout, color, and interaction for low vision people to access sheet music on tablets relying on their residual vision. \\ \cline{3-5}
  &  
    & Weasel: a computer based system for providing non-visual access to music notation
    & \cite{challis2000weasel}
    & This paper presented a multi-modal system that combines a tactile interface and audio output designed to access music notation. \\ \cline{3-5}
  &  
    & Accessibility of Graphic Scores: Design and Exploration of Tactile Supports for Blind People
    & \cite{manenti2024accessibility}
    & Manenti and Ardan presented 3D pinted tactile graphic scores for BLV people to read music and collaborate \\ \cline{2-5}

  % Technology for Learning Music (6)
  & \multirow{6}{=}{Technology for Learning Music}
    & A supporting system of chorus singing for visually impaired persons using depth image sensor
    & \cite{kawarazaki2014supporting}
    & This paper developed an electronic music baton that can detect conductors’ movements and provide the movements’ information to BLV singers via haptic vibrations, tested only with non-BLV conductors. \\ \cline{3-5}
  &  
    & Tactile Presentation of Orchestral Conductor's Motion Trajectory
    & \cite{ueda2024tactile}
    & In this paper, vibrotactile actuators have been used to convey the conductor’s hand and baton movements to musicians, tested with non-BLV participants. \\ \cline{3-5}
  &  
    & Musical Haptic Wearables for Synchronisation of Visually-impaired Performers: a Co-design Approach
    & \cite{turchet2021musical}
    & This paper probed collaboration among BLV individuals via a series of studies to improve synchronization through vibrotactile armbands, belts, and bracelets that provide feedback on tempo, volume and timing. \\ \cline{3-5}
  &  
    & DigituSync: A Dual-User Passive Exoskeleton Glove That Adaptively Shares Hand Gestures
    & \cite{nishida2022digitusync}
    & This paper presented an augmented glove has been designed to aid teachers in conveying correct hand positions by adjusting to the user’s hand movements. \\ \cline{3-5}
  &  
    & NoteBlock: Prototype Design of Music Learning Experience for Blind and Low Vision Children in Preschool Ages
    & \cite{liu2024noteblock}
    & Liu et al. proposed a tactile 3D interface for learning music notes and improving music literacy. \\ \cline{3-5}
  &  
    & A longitudinal study of haptic pitch correction guidance for string instrument players
    & \cite{yoo2017longitudinal}
    & This paper presented vibrotactile actuators on the user’s arms have been used to provide tuning feedback through variations in vibration intensity, reducing the need for visual cues, tested with non-BLV participants \\ \cline{2-5}

  % Technology for Studio Activities (7)
  & \multirow{7}{=}{Technology for Studio Activities}
    & “Composability”: widening participation in music making for people with disabilities via music software and controller solutions
    & \cite{anderson1996composability}
    & The authors presented a system for composing and editing music with simplified mouse clicks and keyboard shortcuts to facilitate access for BLV users. \\ \cline{3-5}
  &  
    & Musica Parlata: a methodology to teach music to blind people
    & \cite{capozzi2012musica}
    & This paper presented an application for producing music, that implemented editing shortcuts directly on a keyboard. \\ \cline{3-5}
  &  
    & Haptic Wave: A Cross-Modal Interface for Visually Impaired Audio Producers
    & \cite{tanaka2016haptic}
    & Tanaka et al. developed a haptic digital audio workstation in a wooden board that employed motorized sliders and an infrared sensor for audio editing via tactile interaction. This is one of the prototypes which tested in a real-world context as producers’ home studio.  \\ \cline{3-5}
  &  
    & Non-visual object generation model to ease music notation script access for visually impaired
    & \cite{ranasinghe2017non}
    & Ranasinghe and Jayaratne developed a synthesized singing model that facilitates composing and editing via non-visual text-based cues.  \\ \cline{3-5}
  &  
    & WELLE - a web-based music environment for the blind
    & \cite{vetter2020welle}
    & Vetter developed a web-based platform that creates and records music through a simplified text-based interface controlled with a computer keyboard. \\ \cline{3-5}
  &  
    & Music editors for visually-impaired persons: user interface specifications and system design
    & \cite{malliopoulos2001music}
    & Malliopoulos et al. designed a system that relies on text-to-speech feedback to facilitate music editing, while also mimicking other software environments to minimize the learning curve. \\ \cline{3-5}
  &  
    & “Give me happy pop songs in C major and with a fast tempo”: A vocal assistant for content-based queries to online music repositories
    & \cite{turchet2023give}
    & Turchet et al. proposed an online platform to search, play and create music. \\ \cline{2-5}

  % New Digital Musical Instruments (8)
  & \multirow{8}{=}{New Digital Musical Instruments}
    & PocoPoco: a tangible device that allows users to play dynamic tactile interaction
    & \cite{kanai2011pocopoco}
    & This paper developed a tactile board for playing loop-based music, similar to a MIDI sequencer. \\ \cline{3-5}
  &  
    & Music learning support system using blocks
    & \cite{kobayashi2022music}
    & This paper presented a set of blocks with different musical functions (start, play notes and rest) as USB music controller. \\ \cline{3-5}
  &  
    & A tangible MIDI sequencer for visually impaired people
    & \cite{haenselmann2009tangible}
    & This paper relied on shortcuts: one designed a piano keyboard equipped with shortcuts to change tonalitie. \\ \cline{3-5}
  &  
    & Non-Visual Beats: Redesigning the Groove Pizza
    & \cite{payne2019non}
    & This paper developed a web-based drum sequencer controlled by shortcuts on a computer keyboard. \\ \cline{3-5}
  &  
    & A music application for visually impaired people using daily goods and stationeries on the table
    & \cite{yairi2012music}
    & This paper introduced three cards used as controllers, where their movements were mapped to piano, guitar, and percussion sounds, and they have collaboration. \\ \cline{3-5}
  &  
    & Collaborative music application for visually impaired people with tangible objects on table
    & \cite{omori2013collaborative}
    & This paper presented a device that allows two individuals to create music by manipulating cards connected by ropes (mapped to seven basic music loops) and a dice (mapped to six different musical instruments).  \\ \cline{3-5}
  &  
    & MusiCane: an Accessible Digital Instrument inspired by the white cane
    & \cite{dimogerontakis2024musicane}
    & Dimogerontakis et al. facilitated the spatial issues with cane, the most familiar tool for BLV people. \\ \cline{3-5}
  &  
    & Acoustic Garden: Exploring Accessibility and Interactive Music with Distance-related Audio Effect Modulation in XR
    & \cite{xie2024acoustic}
    & Xie et al. proposed a musical interface in XR which used the garden the real world scenes. \\ \cline{2-5}

  % Music Technology for Non-musical Skills Acquisition (10)
  & \multirow{10}{=}{Music Technology for Non-musical Skills Acquisition.}
    & Sensory puzzles
    & \cite{lackner1999sensory}
    & A sensory puzzle was developed to provide auditory and tactile feedback during a music creation experience, promoting both creative thinking and social communication through the reorganization of musical patterns. \\ \cline{3-5}
  &  
    & CubeHarmonic: a new musical instrument based on Rubik's cube with embedded motion sensor
    & \cite{mannone2019cubeharmonic}
    & Another study proposed a Rubik’s cube whose faces were mapped to musical notes and chords, enabling users to generate music as a way to stimulate abstract thinking. \\ \cline{3-5}
  &  
    & A Framework to Assess Melodic Effectiveness in Training Computational Thinking to Visually Impaired People
    & \cite{costa2022framework}
    & Costa et al. introduced tangible blocks that can be rearranged to create melodic sequences, supporting the development of computational thinking.  \\ \cline{3-5}
  &  
    & Tangible Music Programming Blocks for Visually Impaired Children
    & \cite{sabuncuoglu2020tangible}
    & This paper presented tangible blocks of varying shapes for composing melodies, aimed at facilitating the under- standing of programming concepts and computational thinking. \\ \cline{3-5}
  &  
    & Blocks4All: overcoming accessibility barriers to blocks programming for children with visual impairments
    & \cite{10.1145/3173574.3173643}
    & Milne and Ladner designed tangible blocks embedded with music to teach programming languages, such as Scratch and Blockly, through block reorganization. \\ \cline{3-5}
  &  
    & Live Coding Ensemble as Accessible Classroom
    & \cite{payne2023live}
    & Live coding alongside music performance is way to understand and enjoy music while gaining computational skills. Payne et al. presented a long-term series of studies in real context at school for cultivating live coding by collaborative music performing \\ \cline{3-5}
  &  
    & Approaches to Making Live Code Accessible in a Mixed-Vision Music Ensemble
    & \cite{payne2023approaches}
    & Payne et al. presented a long-term series of studies in real context at school for cultivating live coding by collaborative music performing \\ \cline{3-5}
  &  
    & "Different and Boundary-Pushing:" How Blind and Low Vision Youth Live Code Together
    & \cite{payne2024different}
    & Payne et al. presented live codin collaborative music performing. \\ \cline{3-5}
  &  
    & Redesigning EarSketch for Inclusive CS Education: A Participatory Design Approach
    & \cite{ding2024redesigning}
    & Ding et al. co-designed with teachers of BLV people a browser-based application for content browsing, music and code editing. \\ \cline{3-5}
  &  
    & Musical Performances in Virtual Reality with Spatial and View-Dependent Audio Descriptions for Blind and Low-Vision Users
    & \cite{dang2024musical}
    & This paper proposed the musical appreciation and learning in virtually reality with AI audio description of instrument and environment settings to help with a immersive musical performance. \\ \cline{2-5}

  % Musical Games and Entertainment (4)
  & \multirow{4}{=}{Musical Games and Entertainment}
    & Virtual Stage: An Immersive Musical Game for People with Visual Impairment
    & \cite{de2015virtual}
    & This paper developed binaural audio techniques to create a 3D audio environment, allowing players to have a better immersive experience while playing musical instruments in the game. \\ \cline{3-5}
  &  
    & TapBeats: accessible and mobile casual gaming
    & \cite{kim2011tapbeats}
    & This paper proposed a game based on musical rhythms for Android. \\ \cline{3-5}
  &  
    & Finger dance: a sound game for blind people
    & \cite{miller2007finger}
    & This paper presented an original audio-based rhythm-action game designed to entertain BLV individuals and facilitate collaboration with non-BLV people.  \\ \cline{3-5}
  &  
    & Blind hero: enabling guitar hero for the visually impaired
    & \cite{yuan2008blind}
    & This paper introduced a haptic glove that provides an alternative approach to playing the existing musical games with tactile stimuli.  \\ \cline{1-5}

% ——— Not Presenting New Technologies (10 行) ———
\multirow{10}{=}{Not Presenting New Technologies}

  % Collecting Insights for Future Design (6)
  & \multirow{6}{=}{Collecting Insights for Future Design}
    & Sensorimotor Synchronization in Blind Musicians: Does Lack of Vision Influencenon-verbal Musical Communication?
    & \cite{volta2021sensorimotor}
    & Music performers in orchestras or groups often rely on non-verbal communication, such as visual cues and eye contact, which can be challenging for BLV players who struggle to perceive these signals \\ \cline{3-5}
  &  
    & "We Musicians Know How to Divide and Conquer": Exploring Multimodal Interactions To Improve Music Reading and Memorization for Blind and Low Vision Learners
    & \cite{lu2024we}
    & Lu et al. explored different multimodal interactions to improve readability of music scores. \\ \cline{3-5}
  &  
    & Playing with Feeling: Exploring Vibrotactile Feedback and Aesthetic Experiences for Developing Haptic Wearables for Blind and Low Vision Music Learning
    & \cite{lu2023playing}
    & Lu et al. investigated the use of vibrotactile patterns and materials to assist BLV people in making music. \\ \cline{3-5}
  &  
    & Towards Accessible Musical Performances in Virtual Reality: Designing a Conceptual Framework for Omnidirectional Audio Descriptions
    & \cite{dang2024towards}
    & Dang et al. investigated the current state of practice and challenges of VR music experience, and giving design implication for the audio descriptions design. \\ \cline{3-5}
  &  
    & Opportunities for Accessible Virtual Reality Design for Immersive Musical Performances for Blind and Low-Vision People
    & \cite{dang2023opportunities}
    & Dang et al. collected data to explore the opportunity of creating VR musical concerts for BLV people. \\ \cline{3-5}
  &  
    & A comparative study of verbal descriptions of emotions induced by music between adults with and without visual impairments
    & \cite{park2020comparative}
    & Park et al. examined differences in emotional perception between BLV and non-BLV individuals, proposing design solutions based on their findings. \\ \cline{2-5}

  % Studying Existing Devices (3)
  & \multirow{3}{=}{Studying Existing Devices}
    & How Blind and Visually Impaired Composers, Producers, and Songwriters Leverage and Adapt Music Technology
    & \cite{payne2020blind}
    & This study focused on existing devices for making music, and interviewed BLV individuals to examine how they use available technology. \\ \cline{3-5}
  &  
    & “Why are there so many steps?”: Improving Access to Blind and Low Vision Music Learning through Personal Adaptations and Future Design Ideas
    & \cite{lu2023there}
    & Lu et al. investigated the current state of practice and challenges in the music learning, and giving insights for the future technology development. \\ \cline{3-5}
  &  
    & Empowering Blind Musicians to Compose and Notate Music with SoundCells
    & \cite{payne2022empowering}
    & Payne et al. tested an accessible piece of music technology in a real-world setting. \\ \cline{2-5}

  % Keynote (1)
  & \multirow{1}{=}{Keynote}
    & Musical Multimodal Interaction: From Bodies to Ecologies
    & \cite{tanaka2020musical}
    & This keynote paper reviewed the advancements in music technology for BLV individuals over the past few years. \\
\end{longtable}
\end{landscape}

\end{document}